\documentclass[a4paper,fleqn,usenatbib]{mnras}

\usepackage{newtxtext,newtxmath}

\usepackage[T1]{fontenc}
\usepackage{ae,aecompl}

\usepackage{graphicx}	
\usepackage{amsmath}	

\newcommand{\ASCA}{{\it ASCA}}
\newcommand{\Chandra}{{\it Chandra}}
\newcommand{\RXTE}{{\it RXTE}}
\newcommand{\Suzaku}{{\it Suzaku}}
\newcommand{\XMM}{{\it XMM-Newton}}
\newcommand{\cxoj}{CXOU~J160103.1$-$513353}
\newcommand{\wgaj}{1WGA~J1713.4$-$3949}
\newcommand{\xmmj}{XMMU~J172054.5$-$372652}
\newcommand{\casa}{CXOU~J232327.9+584842}
\newcommand{\NH}{N_{\rm H}}
\newcommand{\Rem}{R_{\rm em}}
\newcommand{\fabs}{f^{\rm abs}_{\rm 0.5-10}}
\newcommand{\Msun}{M_{\rm Sun}}

\title[X-ray bounds on young CCOs]{X-ray bounds on cooling, composition, and magnetic field of the Cassiopeia~A neutron star and young central compact objects}

\author[W. C. G. Ho et al.]{Wynn C. G. Ho$^{1}$\thanks{E-mail: wynnho@slac.stanford.edu},
Yue Zhao$^{2}$,
Craig O. Heinke$^{2}$,
D. L. Kaplan$^{3}$,
Peter S. Shternin$^{4}$,
and
\newauthor
M. J. P. Wijngaarden$^{5}$
\\
$^{1}$Department of Physics and Astronomy, Haverford College, 370 Lancaster Avenue, Haverford, PA, 19041, USA\\
$^{2}$Department of Physics, University of Alberta, CCIS 4-183, Edmonton, AB T6G 2E1, Canada\\
$^{3}$Center for Gravitation, Cosmology, and Astrophysics, Department of Physics, University of Wisconsin-Milwaukee, P.O. Box 413, Milwaukee, WI 53201, USA\\
$^{4}$Ioffe Institute, Politekhnicheskaya 26, St Petersburg, 194021, Russia\\
$^{5}$Mathematical Sciences and STAG Research Centre, University of Southampton, Southampton SO17 1BJ, UK\\
}

\date{Accepted 2021 July 16. Received 2021 July 16; in original form 2021 March 16}

\pubyear{2021}

\begin{document}
\label{firstpage}
\pagerange{\pageref{firstpage}--\pageref{lastpage}}
\maketitle

\begin{abstract}
We present analysis of multiple \Chandra\ and \XMM\ spectra, separated
by 9--19 years, of four of the youngest central compact objects (CCOs)
with ages $<2500\mbox{ yr}$: \casa\ (Cassiopeia~A), \cxoj\ (G330.2+1.0),
\wgaj\ (G347.3$-$0.5), and \xmmj\ (G350.1$-$0.3).
By fitting these spectra with thermal models, we attempt to constrain
each CCO's long-term cooling rate, composition, and magnetic field.
For the CCO in Cassiopeia~A, 14 measurements over 19 years indicate
a decreasing temperature at a ten-year rate of $2.2\pm0.2$ or
$2.8\pm0.3$~percent ($1\sigma$ error) for a constant or changing X-ray
absorption, respectively.
We obtain cooling rate upper limits of 17~percent for \cxoj\
and 6~percent for \xmmj.
For the oldest CCO, \wgaj, its temperature seems to have increased
by $4\pm2$~percent over a ten year period.
Assuming each CCO's preferred distance and
an emission area that is a large fraction of the total stellar surface,
a non-magnetic carbon atmosphere spectrum is a good fit to spectra
of all four CCOs.
If distances are larger and emission areas are
somewhat smaller, then equally good spectral
fits are obtained using a hydrogen atmosphere with
$B\le7\times 10^{10}\mbox{ G}$ or $B\ge10^{12}\mbox{ G}$ for \cxoj\
and $B\le10^{10}\mbox{ G}$ or $B\ge10^{12}\mbox{ G}$ for \xmmj\
and non-magnetic hydrogen atmosphere for \wgaj.
In a unified picture of CCO evolution, our results suggest most CCOs,
and hence a sizable fraction of young neutron stars, have a surface
magnetic field that is low early in their life but builds up over
several thousand years.
\end{abstract}




\section{Introduction} \label{sec:intro}

Central compact objects (CCOs) form a class of young neutron stars that
are found near the center of a supernova remnant, possess relatively
steady long-term thermal X-ray surface emission, and are not clearly
detected at other energies \citep{deluca08,deluca17,gotthelfetal13}.
Three of the most notable CCOs are those in supernova remnants (SNRs)
1E~1207.4$-$5209, Kesteven~79, and Puppis~A, which are the only three thus
far with a measured spin period ($P=424$, 105, and 112~ms, respectively).
The 340~yr old SNR Cassiopeia~A \citep{fesenetal06} hosts another
important CCO, one that is apparently rapidly decreasing in temperature
at a rate of $\lesssim 3$~percent per decade
\citep{heinkeho10,posseltpavlov18,wijngaardenetal19},
which is naturally explained by and provides direct evidence of
proton superconductivity and neutron superfluidity in the core of
the neutron star \citep{pageetal11,shterninetal11},
although other explanations or contributions are possible (see, e.g.,
\citealt{yangetal11} for r-mode oscillations, \citealt{blaschkeetal12}
for slower thermal relaxation, \citealt{negreirosetal13,tarantoetal16}
for rotation-induced and fast cooling,
\citealt{nodaetal13,sedrakian13} for phase transitions,
\citealt{bonannoetal14} for magnetic field decay, and
\citealt{leinson14,hamaguchietal18} for axion cooling).

\begin{table*}
\centering
\caption{Properties of CCOs studied here.  Note that pulsed fraction is not defined consistently among referenced works.  See text for details and references.}
\label{tab:cco}
\begin{tabular}{cccccc}
\hline
 & & & & \multicolumn{2}{c}{Pulsation search results} \\
CCO & SNR & Age & $d$ & Pulsed fraction & Search range \\
& & (yr) & (kpc) & & \\
\hline
\cxoj & G330.2+1.0 & $<1000$ & 5--11.3 & $<0.40$ & $P>12.5\mbox{ ms}$ \\
 & & & & $<0.21$ & $147\mbox{ ms}<P<100\mbox{ s}$ \\
\wgaj & G347.3$-$0.5 & 1500--2300 & $1.3\pm0.4$ & not given & $3\mbox{ ms}<P<10\mbox{ ms}$ \\
 & & & & $<0.25$ & $7.8\mbox{ ms}<P<100\mbox{ s}$ \\
 & & & & $<0.04$ & $400\mbox{ ms}<P<100\mbox{ s}$ \\
\xmmj & G350.1$-$0.3 & $<600$ & 4.5--10.7 & $<0.33$ & $146\mbox{ ms}<P<1.8\mbox{ s}$ \\
 & & & & $<0.20$ & $1.8\mbox{ s}<P<1\mbox{ hr}$ \\
\casa & Cassiopeia~A & $340\pm19$ & $3.4^{+0.3}_{-0.1}$ & $<0.12$ & $P>10\mbox{ ms}$ \\
\hline
\end{tabular}
\end{table*}

To model and interpret CCO spectra, it is important to know the
composition of the thin atmosphere that covers the neutron star surface
\citep{potekhin14}.
Because of rapid gravitational settling of heavy elements, the surface
is composed of the lightest elements present \citep{alcockillarionov80}.
For isolated or non-accreting binary systems, a hydrogen atmosphere is
likely due to even a small amount of accretion from the interstellar
medium \citep{blaesetal92}.
Helium or carbon atmospheres are possible for the youngest neutron
stars that are still hot after formation since hydrogen and helium can
be depleted by residual nuclear burning
\citep{changbildsten03,changbildsten04,changetal10,wijngaardenetal19,wijngaardenetal20}.
Such appears to be the case for the CCO in Cassiopeia~A,
i.e, its observed properties
are well-described by a carbon atmosphere, the presence of which can be
understood given the youth of the CCO \citep{changetal10,wijngaardenetal19}.
In recent years, a few other CCOs are suggested to also have carbon
atmospheres
\citep{klochkovetal13,klochkovetal15,klochkovetal16,suleimanovetal17,doroshenkoetal18,hebbaretal20,potekhinetal20},
although it is important to note that these CCOs/SNRs are likely to
be much older than Cassiopeia~A.

Another important property of CCOs is their magnetic field $B$, which,
as a class, is uncertain.
From their timing properties, the magnetic fields of the three CCOs with
a measured spin period noted above are inferred to be
$\sim 3\times 10^{10}-10^{11}\mbox{ G}$
\citep{halperngotthelf10,gotthelfetal13},
and that of 1E~1207.4$-$5209 is in approximate agreement with the field
inferred from an electron cyclotron interpretation of features in its observed
spectrum \citep{sanwaletal02,bignamietal03,delucaetal04}.
Meanwhile, non-detection of pulsations from the CCO in Cassiopeia~A
\citep{murrayetal02,ransom02,pavlovluna09,halperngotthelf10}
suggests its observed X-rays are due to emission from the entire
hot stellar surface, which produces a spectrum well-fit by a
low magnetic field carbon atmosphere \citep{hoheinke09}.
The low inferred magnetic field of these CCOs could be the
result of birth at these field strengths.
Alternatively, large and rapid mass accretion soon after a
supernova explosion \citep{chevalier89} can bury an initial surface
field and lead to the field's gradual emergence at the present time
\citep{ho11,ho13}.

Here we examine \Chandra\ and \XMM\ CCD spectra of three CCOs
(see Table~\ref{tab:cco})
that are possibly the next youngest to the CCO in Cassiopeia~A:
\cxoj\
with an age of $<1000\mbox{ yr}$ \citep{borkowskietal18},
\wgaj\
with an age of $\sim1500-2300\mbox{ yr}$ \citep{tsujiuchiyama16,aceroetal17},
and \xmmj\
with an age of $<600\mbox{ yr}$ \citep{borkowskietal20}.
\Chandra\ spectra of each CCO are taken approximately nine to fourteen
years apart, and \XMM\ spectra of \wgaj\ span either
thirteen years for EPIC-MOS or ten years for EPIC-pn (see Table~\ref{tab:data}).
These \Chandra\ ACIS data are all the publically-available data
on each of these CCOs; similarly for \XMM\ data on \wgaj.
We do not examine here two \XMM\ observations of \cxoj\ taken in
2008 and 2015 since they have a shorter time separation (7~years)
than our \Chandra\ data (11~years) and are analyzed in previous
work (see, e.g., \citealt{doroshenkoetal18}).
We also do not examine the single \XMM\ observation of \xmmj\
taken in 2007 (see \citealt{gaensleretal08}).
While the first epoch \Chandra\ CCO spectra considered here are well-studied
(similarly for the first epoch \XMM\ spectra and 2014 pn spectrum of \wgaj;
see references below), the later epoch CCO spectra have not been
previously analyzed (see \citealt{borkowskietal18,borkowskietal20}
for analyses of the spectra of SNRs G330.2+1.0 and G350.1$-$0.3
using the observations examined here).
Therefore our work is the first that we are aware to investigate the
possibility of long-term variability of each of these CCOs.
Using all epochs of data, we attempt to constrain the cooling rate and infer
properties of each CCO, in particular their surface composition and
magnetic field.
We also update \citet{wijngaardenetal19} with spectral measurements
and fits of the CCO in Cassiopeia~A, including the most recent
\Chandra\ observation from 2019 and using the latest improvements
in ACIS calibration,
especially contaminant model N0014, which improves upon N0010 used by
\citet{posseltpavlov18} and N0012 used by \citet{wijngaardenetal19}
(see Section~\ref{sec:data}).

\begin{table}
\centering
\caption{X-ray observations of four CCOs.}
\label{tab:data}
\begin{tabular}{clccc}
\hline
ObsID & Date & Exposure & Frame & Pile-up \\
 & & & time & fraction \\
 & & (ks) & (s) & \\
\hline
\multicolumn{5}{c}{\cxoj\ (\Chandra\ ACIS-I)} \\
 6687 & 2006 May 22 & 50 & 3.24 & 0.02 \\
19163 & 2017 May  2 & 74 & 3.14 & 0.01 \\
20068 & 2017 May  5 & 74 & 3.14 & 0.01 \\
\\
\multicolumn{5}{c}{\wgaj\ (\Chandra\ ACIS-I)} \\
  736 & 2000 July 25 & 30 & 3.24 & 0 \\
5559$^{\rm a}$ & 2005 April 19 & 9.6 & 3.24 & 0.30 \\
15967 & 2014 July 6 & 29 & 0.94 & 0.03 \\
\\
\multicolumn{5}{c}{\wgaj\ (\XMM\ EPIC)} \\
0093670501$^{\rm b}$ & 2001 March 2 & 15 & 2.6/0.2 & 0 \\
0207300201$^{\rm c}$ & 2004 February 22 & 34 & 2.6/0.006 & 0 \\
0722190101$^{\rm d}$ & 2013 August 24 & 139 & 0.002/0.006 & 0 \\
0740830201$^{\rm c}$ & 2014 March 2 & 141 & 2.6/0.006 & e \\
\\
\multicolumn{5}{c}{\xmmj\ (\Chandra\ ACIS-S)} \\
10102 & 2009 May 22 & 83 & 3.24 & 0.06 \\
14806 & 2013 May 11 & 89 & 0.00285 & f \\
20312 & 2018 July 2 & 41 & 3.14 & 0.05 \\
20313 & 2018 July 4 & 20 & 3.14 & 0.05 \\
21120 & 2018 July 5 & 38 & 3.14 & 0.05 \\
21119 & 2018 July 7 & 48 & 3.14 & 0.05 \\
21118 & 2018 July 8 & 43 & 3.14 & 0.05 \\
\\
\multicolumn{5}{c}{Cassiopeia~A (\Chandra\ ACIS-S Graded)} \\
   114 & 2000 Jan 30 & 50 & 3.24 & 0.13 \\
  1952 & 2002 Feb 6  & 50 & 3.24 & 0.13 \\
  5196 & 2004 Feb 8  & 50 & 3.24 & 0.13 \\
  9117 & 2007 Feb 5  & 25 & 3.04 & 0.11 \\
  9773 & 2007 Feb 8  & 25 & 3.04 & 0.12 \\
 10935 & 2009 Nov 2  & 23 & 3.04 & 0.11 \\
 12020 & 2009 Nov 3  & 22 & 3.04 & 0.11 \\
 10936 & 2010 Oct 31 & 32 & 3.04 & 0.11 \\
 13177 & 2010 Nov 2  & 17 & 3.04 & 0.11 \\
 14229 & 2012 May 15 & 49 & 3.04 & 0.08 \\
 14480 & 2013 May 20 & 49 & 3.04 & 0.10 \\
 14481 & 2014 May 12 & 49 & 3.04 & 0.10 \\
 14482 & 2015 Apr 30 & 49 & 3.04 & 0.07 \\
 19903 & 2016 Oct 20 & 25 & 3.04 & 0.08 \\
 18344 & 2016 Oct 21 & 26 & 3.04 & 0.09 \\
 19604 & 2017 May 16 & 50 & 3.04 & 0.07 \\
 19605 & 2018 May 15 & 49 & 3.04 & 0.07 \\
 19606 & 2019 May 13 & 49 & 3.04 & 0.06 \\
\hline
\multicolumn{5}{l}{$^{\rm a}$ Not used here due to 30~percent pile-up fraction. Used by} \\
\multicolumn{5}{l}{\citet{mignanietal08} only to determine position of CCO.} \\
\multicolumn{5}{l}{$^{\rm b}$ MOS/pn in full frame/extended full frame mode; pn  not used here.} \\
\multicolumn{5}{l}{$^{\rm c}$ MOS/pn in full frame/small window mode.} \\
\multicolumn{5}{l}{$^{\rm d}$ MOS/pn in timing/small window mode; MOS not used here.} \\
\multicolumn{5}{l}{$^{\rm e}$ Pile-up fraction of 0.03 for MOS1 and 0 for MOS2 and pn.} \\
\multicolumn{5}{l}{$^{\rm f}$ Not used here due to continuous-clocking mode.}
\end{tabular}
\end{table}

An outline of the paper is as follows.
Section~\ref{sec:ccosummary} provides some details and summarizes relevant
previous works on the four young CCOs examined here. 
Section~\ref{sec:data} describes our procedure for analyzing
\Chandra\ and \XMM\ spectra of the four CCOs: \cxoj, \wgaj, \xmmj,
and Cassiopeia~A.
Section~\ref{sec:results} presents our spectral fits.
Section~\ref{sec:discuss} summarizes and discusses implications
of our results in the context of neutron star cooling and the CCO
class of neutron stars.

\vspace{-2em}
\section{Summary of four CCOs} \label{sec:ccosummary}

The first CCO we consider, \cxoj, is in SNR G330.2+1.0 \citep{parketal06}
at a distance $d>4.9\pm0.3\mbox{ kpc}$ \citep{mccluregriffithsetal01},
with a possible upper limit of 11.3~kpc \citep{toriietal06}, although
\citet{borkowskietal18} argue against $d\gtrsim 10\mbox{ kpc}$.
From measurements of shell expansion, the SNR age is found to be
$<1000$~yr \citep{borkowskietal18}.
Pulsation searches yield pulsed fraction limits of 40~percent for
spin periods $P>12.5\mbox{ ms}$ using 2008 \XMM\ data \citep{parketal09} and
21~percent for $147\mbox{ ms}<P<100\mbox{ s}$ using 2008 and 2015 \XMM\ data
\citep{doroshenkoetal18}.
\citet{doroshenkoetal18}
use a two-component (hot spot and cool entire surface) emission model to
argue that there is a 30~percent probability the viewing geometry is
not favorable for detecting pulsations.
Spectrally, \citet{parketal06} analyzed the early epoch \Chandra\ data
studied here (ObsID~6687),
\citet{parketal09} performed a joint fit with 2008 \XMM\ spectra,
and \citet{doroshenkoetal18} fit only the 2008 \XMM\ spectra along with
subsequent \XMM\ data from 2015 using non-magnetic hydrogen and
carbon atmosphere spectra and did not search for temperature
variations between the 2008 and 2015 data.
While power law and thermal blackbody or atmosphere model spectra
can each fit the observed spectra of \cxoj, the power law model
is disfavored because of a large index $\Gamma\gtrsim 5$ and high
absorption $\NH\approx5.4\times 10^{22}\mbox{ cm$^{-2}$}$ \citep{parketal09}.
The latter can be compared to
$\NH\approx(2-3)\times 10^{22}\mbox{ cm$^{-2}$}$ from spectral
analysis of the SNR using \ASCA, \Chandra, and \XMM\
\citep{toriietal06,parketal09,williamsetal18}.

The second CCO, \wgaj, is in SNR G347.3$-$0.5
(also known as RX~J1713.7$-$3946)
\citep{slaneetal99} at a distance of $\sim 1\mbox{ kpc}$
(see, e.g., \citealt{fukuietal03,kooetal04,moriguchietal05}).
Here we adopt $d=1.3\pm0.4\mbox{ kpc}$ from \citet{cassamchenaietal04}
for simplicity.
While the association of the SNR  with the historical supernova SN~393
is uncertain \citep{wangetal97,fesenetal12}, expansion of the remnant
indicates an age $\sim 1500-2300\mbox{ yr}$ \citep{tsujiuchiyama16,aceroetal17}.
Pulsation searches provide a pulsed fraction limit of 25~percent
for $7.8\mbox{ ms}<P<~100\mbox{ s}$ using \RXTE\ and
a limit of 4~percent for $0.4\mbox{ s}<P<~100\mbox{ s}$ using \XMM\
\citep{lazendicetal03} and
no detection for $3\mbox{ ms}<P<10\mbox{ ms}$ using \XMM\
\citep{papaetal20}.
No optical/IR counterpart is detected \citep{mignanietal08}.
\citet{cassamchenaietal04} fit the earliest epoch \XMM\ MOS and pn spectra
used here (ObsID 0093670501) with either a power law or thermal model,
\citet{lazendicetal03} performed a joint fit with the above \XMM\ spectra and
the early epoch
\Chandra\ spectrum used here (ObsID~736),
and \citet{potekhinetal20} fit only the 2014 \XMM\ pn spectrum
(ObsID 0740830201).
Like in the case of \cxoj, a large inferred power law index
$\Gamma=4.2$ and high absorption $\NH=11\times 10^{21}\mbox{ cm$^{-2}$}$
argue for a thermal model.
The absorption can be compared to $\NH\sim5\times 10^{21}\mbox{ cm$^{-2}$}$
for regions of the SNR near the CCO \citep{cassamchenaietal04,aceroetal09}.

The third CCO, \xmmj, is in SNR G350.1$-$0.3 \citep{gaensleretal08}.
The distance is estimated to be 4.5~kpc from a possible molecular
cloud association, 4.5--10.7~kpc from \ion{H}{i} absorption in the
direction of the SNR \citep{gaensleretal08}, or $\sim 9\mbox{ kpc}$ from
using the interstellar gas density \citep{yasumietal14}.
Free expansion of the SNR gives an age 600--1200~yr \citep{lovchinskyetal11},
while ejecta motion yields an age $<600$~yr \citep{borkowskietal20},
in agreement with that obtained from forward and reverse shock modeling
of the SNR \citep{leahyetal20}.
Pulsation searches using \XMM\ yield pulsed fraction limits of
33~percent for $146\mbox{ ms}<P<~1.8\mbox{ s}$ and
20~percent for $1.8\mbox{ s}<P<~1\mbox{ hr}$ \citep{gaensleretal08}.
Either a power law or blackbody model can fit the CCO spectra from \XMM\
\citep{gaensleretal08} and \Chandra\ (ObsID 10102; \citealt{lovchinskyetal11}).
Like in the other CCOs studied here, the large inferred
$\Gamma\approx 5.5$ disfavors the power law model.

The last CCO we consider is the one in the Cassiopeia~A SNR discussed above
(also known as CXOU~J232327.9+584842; \citealt{tananbaum99};
hereafter we do not necessarily distinguish between the CCO and SNR).
The distance to the SNR is $3.4^{+0.3}_{-0.1}\mbox{ kpc}$ \citep{reedetal95},
and expansion of the remnant gives a supernova explosion date of
$1681\pm19$ \citep{fesenetal06}, such that the SNR age is
$\approx 340\mbox{ yr}$.
The strongest limit on the pulsed fraction of the CCO is 12~percent
for $P>10\mbox{ ms}$ using \Chandra\ HRC \citep{halperngotthelf10}.
In addition to carbon atmosphere model fits to the CCO spectra,
a power law model with $\Gamma\approx 5$ or hydrogen atmosphere models
can also fit the spectra \citep{pavlovluna09,hoheinke09,posseltetal13}.

\vspace{-2em}
\section{Spectral analysis of \Chandra\ and \XMM\ data} \label{sec:data}

A summary of \Chandra\ and \XMM\ observations used in this work
is shown in Table~\ref{tab:data}.
Note that \wgaj\ is far off-axis and on the ACIS-I S1 chip in the 2000
observation (see Figure~1 of \citealt{lazendicetal03}).

We reprocess all \Chandra\ data with \texttt{chandra\_repro} and
Chandra Interactive Analysis of Observations (CIAO) 4.13 and
Calibration Database (CALDB) 4.9.4 \citep{fruscioneetal06}.
Examination of the lightcurve (100~s bins) of each observation
provides no indication of background flares.
Spectra are extracted using \texttt{specextract}.
For \cxoj, source counts are taken from a $\approx$2.\!\!\arcsec5
radius circle
centered very near the CCO position determined by \citet{parketal06}
using ObsID~6687, i.e., $\mbox{R.A.}=16^{\rm h}01^{\rm m}03.\!\!^{\rm s}14$,
$\mbox{decl.}=-51^\circ33\arcmin53.\!\!\arcsec6$ (J2000),
while background for the source is taken from a 3\arcsec--5\arcsec\
radius circular annulus around the CCO.
The resulting background-subtracted source counts are $\sim 640$,
710, and 720~counts for ObsIDs~6687, 19163, and 20068, respectively.
Since ObsIDs~19163 and 20068 are taken within a few days of each other,
we use \texttt{combine\_spectra} and \texttt{dmgroup} to merge spectra
extracted from these two observations for a total exposure of 148~ks.
Spectra are binned with a minimum of 25 counts per energy bin.
Using \texttt{pileup\_map}, we find a pile-up fraction of $<2$~percent
(see Table~\ref{tab:data}).

For \Chandra\ data of \wgaj, source counts are taken from a region centered
at the CCO position determined by \citet{mignanietal08} using ObsID~5559,
i.e., $\mbox{R.A.}=17^{\rm h}13^{\rm m}28.\!\!^{\rm s}32$,
$\mbox{decl.}=-39^\circ49\arcmin53.\!\!\arcsec34$ (J2000).
For ObsID~736, the CCO is very far off-axis and is extremely distended.
Therefore, we use a 37\arcsec$\times$61\arcsec\ radius elliptical
region tilted at 30$^\circ$ to extract source counts
and a 2\arcmin$\times$3\arcmin\ radius elliptical
region tilted at 30$^\circ$ and about 4\arcmin\ north of the CCO
for the background;
these regions are similar to those used in \citet{lazendicetal03}.
For ObsID~15967, we use a 6\arcsec$\times$10\arcsec\ radius elliptical
region tilted at 120$^\circ$ to extract source counts
and a 11\arcsec--20\arcsec\ radius circular annulus around the CCO
for the background.
The resulting background-subtracted source counts are $\sim 11000$~counts
each for ObsID~736 and 15967.
Spectra are binned with a minimum of 50 counts per energy bin.
We find negligible pile-up for ObsID~736 and a pile-up fraction of
$<3$~percent for ObsID~15967.

For \XMM\ observations of \wgaj,
we only consider EPIC spectra measured using the same detector mode,
i.e., full frame for MOS and small window for pn (see Table~\ref{tab:data}),
in order to reduce systematic uncertainties.
We use \XMM\ \textsc{science analysis software} (\textsc{sas})
19.0.0 to reduce observation data files (ODFs) for extraction
of spectra.
We process EPIC-MOS and pn ODFs using \texttt{emproc} and
\texttt{epproc}, respectively, to generate calibrated and
concatenated event files.
To clean these files of flaring events, we first
extract high-energy (10--12~keV) and single-event (PATTERN=0) light curves,
from which we determine count rate upper limits for low and steady
background and MOS and pn count-rates of
0.13 counts~s$^{-1}$ and 0.25 counts~s$^{-1}$ for the 2001 dataset,
0.28 counts~s$^{-1}$ and 0.07 counts~s$^{-1}$ for the 2004 dataset,
0.13 counts~s$^{-1}$ and 0.05 counts~s$^{-1}$ for the 2013 dataset, and
0.15 counts~s$^{-1}$ and 0.02 counts~s$^{-1}$ for the 2014 dataset.
These count rates are then input into the \texttt{tabgtigen} tool
to generate corresponding good time intervals (GTIs), which are
then input into the \texttt{evtselect} tool to produce
flare-cleaned event files.
From the cleaned event files, we extract MOS and pn spectra using
a 30\arcsec\ circular region centred on the above-mentioned
source position.
For all MOS full-frame observations, we select for background a
30\arcsec\ circular region centered at
$\mbox{R.A.}=17^{\rm h}13^{\rm m}13^{\rm s}$,
$\mbox{decl.}=-39^\circ47\arcmin58\arcsec$
(J2000), about 3\farcm5 northwest of the source.
In the 2004 and 2014 observations, the source is close to the bottom edge of
the pn field-of-view, and we choose a circular background region centered at
$\mbox{R.A.}=17^{\rm h}13^{\rm m}36^{\rm s}$,
$\mbox{decl.}=-39^\circ47\arcmin56\arcsec$,
about 2\farcm5 northeast of the source.
In the 2013 observation, the source is close to the top edge of the pn
field-of-view, and we select a different location at
$\mbox{R.A.}=17^{\rm h}13^{\rm m}21^{\rm s}$,
$\mbox{decl.}=-39^\circ51\arcmin54\arcsec$ for
background extraction, but it has the same offset (2\farcm5) from the source.
Since non-thermal X-ray emission from the SNR is at a relatively low level
compared to the CCO emission \citep{katsudaetal15}, we find a $<1$~percent
effect on the final results by using different background regions.
Spectra are binned with a minimum of 30 counts per energy bin
for the shorter 2001 and 2004 observations
and 100 counts per energy bin for the longer 2013 and 2014 observations.
Data from MOS1, MOS2, and pn have negligible pile-up, except for the
2014 MOS1 data which have an estimated pile-up fraction of $<3$~percent.

For \Chandra\ data of \xmmj, source counts are taken from a
$\approx$2.\!\!\arcsec5 radius circle
centered very near the CCO position determined by \citet{gaensleretal08}
using \XMM, i.e., $\mbox{R.A.}=17^{\rm h}20^{\rm m}54.\!\!^{\rm s}5$,
$\mbox{decl.}=-37^\circ26\arcmin52\arcsec$ (J2000),
while background for the source is taken from a 3\arcsec--5\arcsec\
radius circular annulus around the CCO.
The resulting background-subtracted source counts are $\sim 3800$,
1700, 830, 1600, 2000, and 1900~counts for ObsIDs~10102, 20312, 20313,
21120, 21119, and 21118, respectively.
Since ObsIDs~20312-3 and 21118-20 are all taken within a week of each other,
we merge
spectra extracted from these observations for a total exposure of 189~ks.
Spectra are binned with a minimum of 50 counts per energy bin.
We find a pile-up fraction of 5--6~percent (see Table~\ref{tab:data}).

For \Chandra\ data of the CCO in Cassiopeia~A,
source counts are taken from a $\approx$2\arcsec\ radius circle
centered on the CCO, while background for the source is taken from a
2.\!\!\arcsec5--4\arcsec\ radius circular annulus around the CCO.
We merge spectra from observations taken within a few days of each other
(i.e., ObsIDs 9117 and 9773, 10935 and 12020, 10936 and 13177, and
19903 and 18344).
Spectra are binned with a minimum of 25 counts per energy bin.
Table~\ref{tab:data} shows the estimated pile-up fraction for each observation.
We note that \texttt{CALDB}~4.9.4 includes the latest ACIS contaminant
model N0014, which corrects for the decreasing quantum efficiency of
ACIS at low
energies\footnote{\url{https://cxc.harvard.edu/caldb/downloads/Release_notes/CALDB_v4.9.4.html}}.

We perform spectral fitting using Xspec 12.11.0 \citep{arnaud96}.
We account for pile-up \citep{davis01} in \Chandra\ data by allowing the grade migration
parameter $\alpha$ to be fit for each observation (unless specified
differently below) and setting the
frame time for each observation to that given in Table~\ref{tab:data},
maximum number of photons to 5, and point-spread-function fraction to 0.95.
As noted in Section~\ref{sec:intro}, a power law could be used to
model spectra of each CCO, but the implied index of $\Gamma>4$
in each case indicates thermal models are more appropriate.
Thus, in the present work, we only consider absorbed thermal models
consisting of \texttt{tbabs} and either \texttt{bbodyrad},
\texttt{nsatmos}, \texttt{nsx}, or \texttt{nsmaxg}.
The former models photoelectric absorption in the interstellar medium,
and we use abundances from \citet{wilmsetal00} and cross-sections from
\citet{verneretal96}.
The thermal component models intrinsic emission from the CCO as either
a blackbody or neutron star atmosphere spectrum.
The simplest model is a blackbody (\texttt{bbodyrad}), and we give
its corresponding fit results primarily as a comparison to results
from previous works.

\begin{table}
\centering
\caption{Neutron star atmosphere models used for spectral fits in Section~\ref{sec:results}.  See text for other models considered.}
\label{tab:atm}
\begin{tabular}{ccc}
\hline
Model & Composition & Magnetic field \\
\hline
\texttt{nsatmos} & H & non-magnetic \\
\texttt{nsx} & C & non-magnetic \\
\texttt{nsmaxg} & H & $10^{10}-3\times10^{13}\mbox{ G}$ \\
\hline
\end{tabular}
\vspace{-1em}
\end{table}

Thermal emission from a neutron star is more correctly described
by models that account for the properties of the atmosphere that
covers the stellar surface, in particular the magnetic field and
composition of the atmosphere \citep{potekhin14}.
We consider atmosphere models that are either non-magnetic
(applicable when $B\lesssim 10^9\mbox{ G}$) or magnetic (with a
single $B$ in the range $10^{10}\mbox{ G}\le B\le 3\times 10^{13}\mbox{ G}$),
and these are listed in Table~\ref{tab:atm}.
The four spectral models with varying magnetic field and surface
temperature (via \texttt{nsmaxg}; see \citealt{hoetal08})
do not fit the data well and will not be discussed further.
For composition, partially ionized carbon, oxygen, and neon spectral
models at $10^{12}$ and $10^{13}\mbox{ G}$ \citep{moriho07} fit the
data poorly, so will also not be discussed further.  On the other hand,
non-magnetic hydrogen (\texttt{nsatmos}; \citealt{heinkeetal06})
and carbon (\texttt{nsx}; \citealt{hoheinke09}) spectral models fit the data
well, as do partially ionized hydrogen models at
a number of single magnetic field values
(\texttt{nsmaxg}; \citealt{hoetal08,ho14,potekhinetal14}).
We note that, while non-magnetic helium (via \texttt{nsx}) yields
fits that are comparable to those of non-magnetic hydrogen in the
cases of \cxoj\ and \xmmj\ and are poorer fits in the case of \wgaj,
we do not give their results in the following sections
since a helium atmosphere is unlikely to be present for isolated
neutron stars at the age and temperature of these CCOs
\citep{changetal10,wijngaardenetal19}.

For each absorbed thermal atmosphere model, as well as absorbed
blackbody model, there are three model fit parameters:
X-ray absorption $\NH$, temperature $T$, and normalization $\Rem/d$
(where applicable, we normalize the emission radius $\Rem$ to a
neutron star radius $R=12\mbox{ km}$ and normalize the distance $d$
to the nominal distance for each CCO).
Each fit parameter can be fit jointly between all observations,
so that there is only one best-fit value for each model and source,
or fit individually for each observation of a source.
For each CCO studied here, except Cassiopeia~A, we first perform
spectral fits with all three parameters jointly fit to all data,
and the fit statistic $\chi^2$ is computed for each fit.
This is done for each model shown in Table~\ref{tab:atm}:
non-magnetic hydrogen (\texttt{nsatmos}) and carbon (\texttt{nsx})
and magnetic hydrogen (\texttt{nsmaxg}).  For \texttt{nsmaxg},
models available in Xspec are at single magnetic field strengths of
$B=10^{10}, 4\times10^{10}, 7\times10^{10}, 10^{11}, 10^{12},
2\times10^{12}, 4\times10^{12}, 7\times10^{12}, 10^{13},
2\times10^{13}, 3\times10^{13}\mbox{ G}$;
we perform fits with each one of these \texttt{nsmaxg} models but
only show results for the one or two with the lowest $\chi^2$.
Next, we perform fits which allow only one of the three fit parameters
to vary between each observation, and we cycle through each
parameter, i.e., first $\NH$ is allowed to vary while $T$ and
$\Rem/d$ are the same between observations, then $T$ is allowed
to vary while $\NH$ and $\Rem/d$ are the same, and finally
$\Rem/d$ is allowed to vary while $\NH$ and $T$ are the same.
We use the F-test to compare the resulting $\chi^2$ of the fits
to determine if allowing one fit parameter to vary improves the fit
sufficiently to justify the decrease in number of degrees of freedom (dof).
Then we perform fits which allow two of the three fit parameters
to vary between each observation, and we cycle through each pair
of parameters and use the F-test to evaluate if the improvement
is significant.
Finally, we perform a fit which allows all three fit parameters
to vary and evaluate if the improvement is significant.
Results shown in the tables in Section~\ref{sec:results} are those
where the F-test justifies the number of independent fit parameters.

For modeling the spectra of the CCO in Cassiopeia~A,
we consider only a carbon atmosphere model, since other models are
studied in previous works \citep{pavlovluna09,hoheinke09,posseltetal13},
and we fix the distance to 3.4~kpc and normalization to 1
(i.e., detected X-rays are from entire neutron star surface).
We also include a component \texttt{spexpcut},
with exponent index $=-2$ and characteristic energy
$E_{\rm cut}(\mbox{keV})=[0.49\,\NH(10^{22}\mbox{ cm$^{-2}$})]^{1/2}$,
to model interstellar scattering \citep{predehletal03},
as done in our previous works; however its inclusion has a minimal
effect on the fits at the spectral resolution of the Cassiopeia~A data
(see, e.g., \citealt{posseltetal13}).
Because of the last, we do not apply this scattering model to
analyses of the other CCO spectra.

\vspace{-2em}
\section{Results} \label{sec:results}

\subsection{\cxoj} \label{sec:cxoj}

\begin{figure}
\begin{center}
\includegraphics[width=0.95\columnwidth]{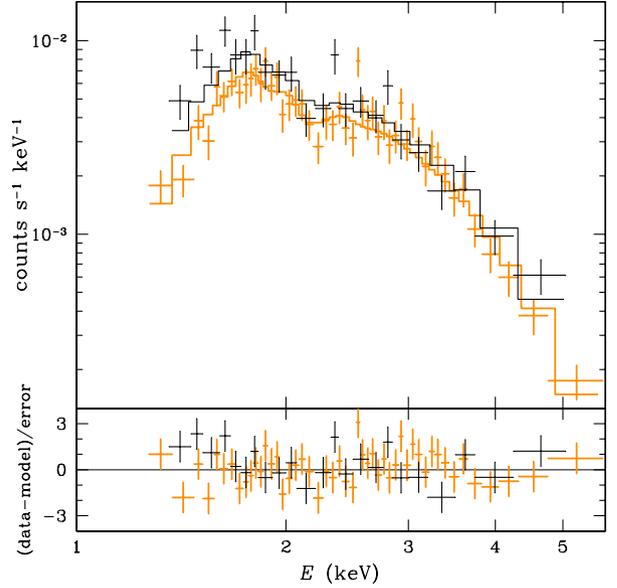}
\caption{
Upper panel shows \Chandra\ spectra (black for 2006 and orange for merged 2017
data) of \cxoj\ and absorbed non-magnetic carbon atmosphere model spectra.
Model results shown here are from fits with all parameters linked between
the two sets of observations.  Lower panel shows fit residuals.
}
\label{fig:spectra_cxoj}
\end{center}
\vspace{-2em}
\end{figure}

\begin{table}
\centering
\caption{Results of fits to the 2006 and (merged) 2017 \Chandra\ spectra of
\cxoj\ at 1.2--6~keV.
Along with \texttt{pileup} and \texttt{tbabs}, fits are performed using
model \texttt{bbodyrad}, \texttt{nsatmos}, \texttt{nsx}, or \texttt{nsmaxg}
with model parameters linked between 2006 and 2017 fits.
Other model parameters are fixed at $M=1.4\,M_\odot$, $R=12\mbox{ km}$,
and $d=5\mbox{ kpc}$.
Errors are 1$\sigma$.}
\label{tab:spectra_cxoj}
\begin{tabular}{ccccc}
\hline
$\NH$ & $T$ & $\Rem/R$ & $f^{\rm abs}_{1.2-6}$ & $\chi^2$/dof \\
 ($10^{22}\mbox{ cm$^{-2}$}$) & ($10^6\mbox{ K}$) & or $\mbox{5 kpc}/d$ & ($10^{-13}\mbox{ erg cm$^{-2}$ s$^{-1}$}$) & \\
\hline
\multicolumn{5}{c}{blackbody} \\
$3.8^{+0.3}_{-0.3}$ & $5.4^{+0.2}_{-0.2}$ & $0.044^{+0.005}_{-0.004}$ & $1.4^{+0.1}_{-0.2}$ & 82.9/67 \\
\\
\multicolumn{5}{c}{\texttt{nsatmos} - non-magnetic hydrogen} \\
$4.6^{+0.3}_{-0.3}$ & $3.0^{+0.2}_{-0.1}$ & $0.23^{+0.05}_{-0.03}$ & $1.4^{+0.1}_{-0.2}$ & 83.1/67 \\
\\
\multicolumn{5}{c}{\texttt{nsx} - non-magnetic carbon} \\
$4.8^{+0.3}_{-0.3}$ & $1.9^{+0.2}_{-0.1}$ & $0.75^{+0.23}_{-0.17}$ & $1.4^{+0.1}_{-0.2}$ & 84.2/67 \\
\\
\multicolumn{5}{c}{\texttt{nsmaxg} - hydrogen at $4\times 10^{10}\mbox{ G}$} \\
$4.5^{+0.3}_{-0.3}$ & $3.1^{+0.2}_{-0.2}$ & $0.20^{+0.04}_{-0.03}$ & $1.4^{+0.1}_{-0.2}$ & 84.9/67 \\
\\
\multicolumn{5}{c}{\texttt{nsmaxg} - hydrogen at $10^{13}\mbox{ G}$} \\
$4.5^{+0.3}_{-0.3}$ & $3.2^{+0.2}_{-0.2}$ & $0.21^{+0.05}_{-0.03}$ & $1.4^{+0.1}_{-0.3}$ & 83.2/67 \\
\hline
\end{tabular}
\end{table}

Figure~\ref{fig:spectra_cxoj} shows the 2006 and (merged) 2017 \Chandra\
spectra of \cxoj, and
Table~\ref{tab:spectra_cxoj} shows results of model fits of these spectra
in the energy range 1.2--6~keV where there are sufficient numbers of counts.
Even though we find the pile-up grade migration parameter $\alpha$
to be unconstrained in all fits of the spectra, we leave it free
to vary since it has an (small) effect on the uncertainties of other
fit parameters.
We see that when we assume all other spectral model parameters
($\NH$, $T$, and $\Rem/d$) are linked and thus constant
between 2006 and 2017, the resulting emission region size $\Rem$
for the blackbody fit is only 0.5~km
(assuming a distance $d=5\mbox{ kpc}$ to \cxoj; see below).
Such a small emitting region or hot spot would naturally lead
to pulsed emission as the neutron star rotates, which is not
seen (see Section~\ref{sec:intro}).  Of course non-detection
could be the result of an unfavorable viewing geometry or
a pulsed fraction below detection thresholds ($\approx 20$~percent
in the case of \cxoj\ and spin periods comparable to other CCOs;
see Table~\ref{tab:cco}).

Very similar quality of fits are achieved using atmosphere models,
but in this case, the emission size is much larger
($\Rem\approx 2-3\mbox{ km}$ for hydrogen and $\sim 9\mbox{ km}$
for carbon, assuming a neutron star radius $R=12\mbox{ km}$)
and indicates a significant fraction of the neutron star surface is
hot and emitting detectable X-rays.
Such large emitting regions can yield low enough amplitude
rotational variations to explain non-detection of pulsations
(see Section~\ref{sec:discuss}).
From similar quality of fits, with a difference in fit-statistic
$\Delta\chi^2\approx 1$ for 67 dof, we cannot
distinguish between hydrogen and carbon compositions.
The same is true for partially ionized hydrogen at magnetic fields
$B\le 7\times 10^{10}\mbox{ G}$ and $B\ge 10^{12}\mbox{ G}$
(see examples in Table~\ref{tab:spectra_cxoj}).
Our results are in general agreement with those of previous fits
to the 2006 \Chandra\ spectrum \citep{parketal06} and \XMM\ spectra
\citep{parketal09,doroshenkoetal18} using a blackbody model,
non-magnetic hydrogen atmosphere (\texttt{hatm})
and carbon atmosphere (\texttt{carbatm}) models, and
fully ionized hydrogen atmosphere model (\texttt{nsa}) at
$B=0, 10^{12}, 10^{13}\mbox{ G}$
(note that the comparison is to the hotter component in two-component
hydrogen model fits performed in \citealt{parketal09,doroshenkoetal18}).
We also note that the absorption
$\NH\sim(4.5-4.8)\times10^{22}\mbox{ cm$^{-2}$}$ inferred for the CCO
using atmosphere models is slightly lower than the power law value of
$(5.4\pm0.6)\times10^{22}\mbox{ cm$^{-2}$}$ \citep{parketal09}
but still greater than the value
$\NH\approx(2-3)\times10^{22}\mbox{ cm$^{-2}$}$ inferred from the
SNR (see Section~\ref{sec:intro}).

There are two important issues to note regarding the spectral fit
results shown in Table~\ref{tab:spectra_cxoj}.
The first is that it is not strictly correct to allow only a fraction
of the neutron star surface to be fit in Xspec using the available
neutron star atmosphere spectral tables, i.e., allowing $\Rem<R$.
This is because these tables are calculated assuming emission from the
entire surface, whereas emission from a hot spot reaches an observer
only for certain angles of photon propagation, which in turn depend
on the location of the spot on the surface and the viewing geometry
and is not a simple re-scaling of emission from the whole surface
(see, e.g., \citealt{bogdanovetal19}).
However, this inaccuracy will be small when the inferred hot spot is large.
The second issue is that each fit assumes the nominal distance to
\cxoj\ of 5~kpc, while the distance could be as large as 11.3~kpc
(see Section~\ref{sec:intro}), although such a large distance would
imply very fast moving supernova ejecta \citep{borkowskietal18}.
The distance assumption directly impacts the inferred size of the
emission region, i.e., $\Rem\propto d$.
One alternative way of interpreting the spectral fit results then is
to increase the distance by a factor of $R/\Rem$ and
assume emission from the entire surface (or a large fraction of it).
This would imply a distance to \cxoj\ of $\sim 22\pm3\mbox{ kpc}$ for a
hydrogen atmosphere or $7\pm2\mbox{ kpc}$ for a carbon atmosphere.
Although the former exceeds the estimated maximum of 11.3~kpc,
this maximum distance can be satisfied if the hydrogen hot spot occupies
$\sim$20~percent of the surface instead of the entire surface.

To complete our spectral analysis of \cxoj, we consider whether
model parameters change between the two epochs of \Chandra\
observations (see Section~\ref{sec:data}).
The greatest improvement in fit statistics occurs when $\NH$ is
allowed to vary, with $\Delta\chi^2\approx6$ for one fewer degree
of freedom and a F-test probability of 0.02--0.03,
and fits only improve by $\Delta\chi^2<1$ when $T$ varies, with
$T$ differences within errors (see Section~\ref{sec:cco}).
When all parameters are allowed to vary, $\Delta\chi^2<12$
and most of the improvement is due to a possible change of $\NH$.
A F-test between allowing only $\NH$ to vary and all parameters
to vary has probability of 0.11.
Thus we find $\NH$ may be varying, but the evidence is not conclusive.

\subsection{\wgaj} \label{sec:wgaj}

\begin{figure}
\begin{center}
\includegraphics[width=0.95\columnwidth]{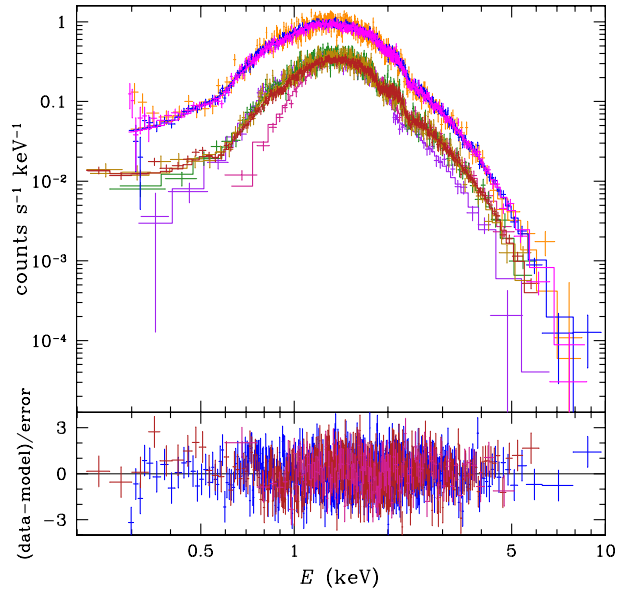}
\caption{
Upper panel shows \Chandra\ ACIS spectra and \XMM\ MOS and pn spectra 
of \wgaj\ and absorbed non-magnetic carbon atmosphere model spectra
(top three are pn, middle three are MOS, and bottom two are \Chandra\ spectra).
Model results shown here are from fits performed independently on
each detector's set of observations and with linked $T$ and
varying $\NH$ and $\Rem$ between observations (for \Chandra) and
with linked $\NH$ and varying $T$ and $\Rem$ between observations (for \XMM).
Lower panel shows fit residuals.
For clarity, only residuals for the 2014 ACIS, 2014 MOS, and 2013 pn
observations are shown; other residuals are similar.
}
\label{fig:spectra_wgaj}
\end{center}
\vspace{-2em}
\end{figure}

\begin{table}
\centering
\caption{Results of fits to the 2001, 2004, and 2014 \XMM\ EPIC-MOS spectra
of \wgaj\ at 0.2--7~keV.
Along with \texttt{tbabs}, fits are performed using
model \texttt{bbodyrad}, \texttt{nsatmos}, \texttt{nsx}, or \texttt{nsmaxg}
with model parameters free to vary between fits, unless given otherwise.
Other model parameters are fixed at $M=1.4\,M_\odot$, $R=12\mbox{ km}$,
and $d=1.3\mbox{ kpc}$, unless otherwise noted.
Absorbed 0.5--10~keV flux $\fabs$ is in $10^{-13}\mbox{ erg cm$^{-2}$ s$^{-1}$}$.
Errors are 1$\sigma$.}
\begin{tabular}{cccccc}
\hline
Year & $\NH$ & $T$ & $\Rem/R$ & $\fabs$ & $\chi^2$/dof \\
& ($10^{21}\mbox{ cm$^{-2}$}$) & ($10^6\mbox{ K}$) & or $\mbox{1.3 kpc}/d$ & & \\
\hline
\multicolumn{6}{c}{blackbody} \\
2001 & $4.5^{+0.2}_{-0.2}$ & $4.75^{+0.05}_{-0.05}$ & $0.044^{+0.001}_{-0.001}$ & $32.0^{+0.3}_{-0.4}$ & $1539/976$ \\
2004 & $4.4^{+0.2}_{-0.2}$ & $4.78^{+0.05}_{-0.05}$ & $0.043^{+0.001}_{-0.001}$ & $30.6^{+0.2}_{-0.4}$ &  \\
2014 & $4.5^{+0.1}_{-0.1}$ & $4.94^{+0.02}_{-0.02}$ & $0.039^{+0.001}_{-0.001}$ & $30.6^{+0.1}_{-0.2}$ &  \\
\\
\multicolumn{6}{c}{\texttt{nsatmos} - non-magnetic hydrogen} \\
2001 & $6.02^{+0.09}_{-0.06}$ & $2.93^{+0.03}_{-0.04}$ & $0.161^{+0.005}_{-0.003}$ & $32.6^{+0.3}_{-0.4}$ & $1086/978$\\
2004 & ---                    & $2.92^{+0.03}_{-0.04}$ & $0.159^{+0.005}_{-0.003}$ & $30.9^{+0.2}_{-0.3}$ & \\
2014 & ---                    & $3.08^{+0.02}_{-0.02}$ & $0.139^{+0.003}_{-0.002}$ & $31.0^{+0.1}_{-0.1}$ & \\
\\
\multicolumn{6}{c}{\texttt{nsx} - non-magnetic carbon} \\
2001 & $7.05^{+0.08}_{-0.06}$ & $1.90^{+0.03}_{-0.04}$ & $0.497^{+0.031}_{-0.021}$ & $32.5^{+0.3}_{-0.4}$ & $1062/978$ \\
2004 & ---                    & $1.88^{+0.03}_{-0.04}$ & $0.492^{+0.028}_{-0.020}$ & $30.9^{+0.2}_{-0.4}$ & \\
2014 & ---                    & $2.05^{+0.02}_{-0.02}$ & $0.389^{+0.011}_{-0.008}$ & $31.0^{+0.1}_{-0.1}$ & \\
\\
\multicolumn{6}{c}{\texttt{nsmaxg} - hydrogen at $10^{10}\mbox{ G}$ ($R=10\mbox{ km}$)} \\
2001 & $6.28^{+0.07}_{-0.08}$ & $3.20^{+0.03}_{-0.03}$ & $0.174^{+0.004}_{-0.004}$ & $32.6^{+0.3}_{-0.5}$ & $1089/978$ \\
2004 & ---                    & $3.17^{+0.03}_{-0.03}$ & $0.172^{+0.004}_{-0.004}$ & $30.9^{+0.2}_{-0.3}$ & \\
2014 & ---                    & $3.34^{+0.02}_{-0.02}$ & $0.154^{+0.002}_{-0.002}$ & $30.9^{+0.1}_{-0.1}$ & \\
\\
\multicolumn{6}{c}{\texttt{nsmaxg} - hydrogen at $3\times10^{13}\mbox{ G}$ ($R=10\mbox{ km}$)} \\
2001 & $6.82^{+0.09}_{-0.08}$ & $3.07^{+0.03}_{-0.04}$ & $0.200^{+0.006}_{-0.005}$ & $32.6^{+0.2}_{-0.4}$ & $1082/978$ \\
2004 & ---                    & $3.06^{+0.03}_{-0.03}$ & $0.197^{+0.006}_{-0.005}$ & $30.9^{+0.3}_{-0.4}$ & \\
2014 & ---                    & $3.22^{+0.02}_{-0.02}$ & $0.172^{+0.003}_{-0.003}$ & $31.0^{+0.1}_{-0.2}$ & \\
\hline
\end{tabular}
\label{tab:xmmmos}
\vspace{-1em}
\end{table}

\begin{table}
\centering
\caption{Results of fits to the 2004, 2013, and 2014 \XMM\ EPIC-pn spectra
of \wgaj\ at 0.3--10~keV.
Along with \texttt{tbabs}, fits are performed using
model \texttt{bbodyrad}, \texttt{nsatmos}, \texttt{nsx}, or \texttt{nsmaxg}
with model parameters free to vary between fits, unless given otherwise.
Other model parameters are fixed at $M=1.4\,M_\odot$, $R=12\mbox{ km}$,
and $d=1.3\mbox{ kpc}$, unless otherwise noted.
Absorbed 0.5--10~keV flux $\fabs$ is in $10^{-13}\mbox{ erg cm$^{-2}$ s$^{-1}$}$.
Errors are 1$\sigma$.}
\label{tab:xmmpn}
\begin{tabular}{cccccc}
\hline
Year & $\NH$ & $T$ & $\Rem/R$ & $\fabs$ & $\chi^2$/dof \\
& ($10^{21}\mbox{ cm$^{-2}$}$) & ($10^6\mbox{ K}$) & or $\mbox{1.3 kpc}/d$ & & \\
\hline
\multicolumn{6}{c}{blackbody} \\
2004 & $4.72^{+0.15}_{-0.15}$ & $4.70^{+0.05}_{-0.05}$ & $0.045^{+0.001}_{-0.001}$ & $30.4^{+0.2}_{-0.3}$ & $1873/1208$ \\
2013 & $4.69^{+0.05}_{-0.05}$ & $4.82^{+0.02}_{-0.02}$ & $0.041^{+0.001}_{-0.001}$ & $29.7^{+0.1}_{-0.1}$ &  \\
2014 & $4.65^{+0.06}_{-0.06}$ & $4.86^{+0.02}_{-0.02}$ & $0.041^{+0.001}_{-0.001}$ & $30.3^{+0.1}_{-0.1}$ &  \\
\\
\multicolumn{6}{c}{\texttt{nsatmos} - non-magnetic hydrogen} \\
2004 & $6.02^{+0.04}_{-0.05}$ & $2.93^{+0.03}_{-0.03}$ & $0.157^{+0.004}_{-0.003}$ & $31.1^{+0.2}_{-0.3}$ & $1303/1210$\\
2013 & ---                    & $3.01^{+0.01}_{-0.01}$ & $0.146^{+0.002}_{-0.002}$ & $30.2^{+0.1}_{-0.1}$ & \\
2014 & ---                    & $3.03^{+0.02}_{-0.01}$ & $0.144^{+0.002}_{-0.002}$ & $30.7^{+0.1}_{-0.1}$ & \\
\\
\multicolumn{6}{c}{\texttt{nsx} - non-magnetic carbon} \\
2004 & $7.06^{+0.04}_{-0.04}$ & $1.90^{+0.03}_{-0.03}$ & $0.485^{+0.025}_{-0.020}$ & $31.1^{+0.1}_{-0.4}$ & $1285/1210$ \\
2013 & ---                    & $1.98^{+0.01}_{-0.01}$ & $0.425^{+0.009}_{-0.009}$ & $30.2^{+0.1}_{-0.2}$ & \\
2014 & ---                    & $2.00^{+0.02}_{-0.02}$ & $0.413^{+0.010}_{-0.009}$ & $30.7^{+0.2}_{-0.2}$ & \\
\\
\multicolumn{6}{c}{\texttt{nsmaxg} - hydrogen at $3\times10^{13}\mbox{ G}$ ($R=10\mbox{ km}$)} \\
2004 & $6.70^{+0.04}_{-0.07}$ & $3.10^{+0.03}_{-0.03}$ & $0.190^{+0.004}_{-0.005}$ & $31.2^{+0.2}_{-0.4}$ & $1368/1210$ \\
2013 & ---                    & $3.17^{+0.02}_{-0.01}$ & $0.176^{+0.001}_{-0.003}$ & $30.2^{+0.2}_{-0.1}$ & \\
2014 & ---                    & $3.20^{+0.02}_{-0.01}$ & $0.174^{+0.001}_{-0.003}$ & $30.8^{+0.1}_{-0.2}$ & \\
\hline
\end{tabular}
\vspace{-1em}
\end{table}

\begin{figure}
\begin{center}
\includegraphics[width=0.45\textwidth,trim=0 10 0 0,clip]{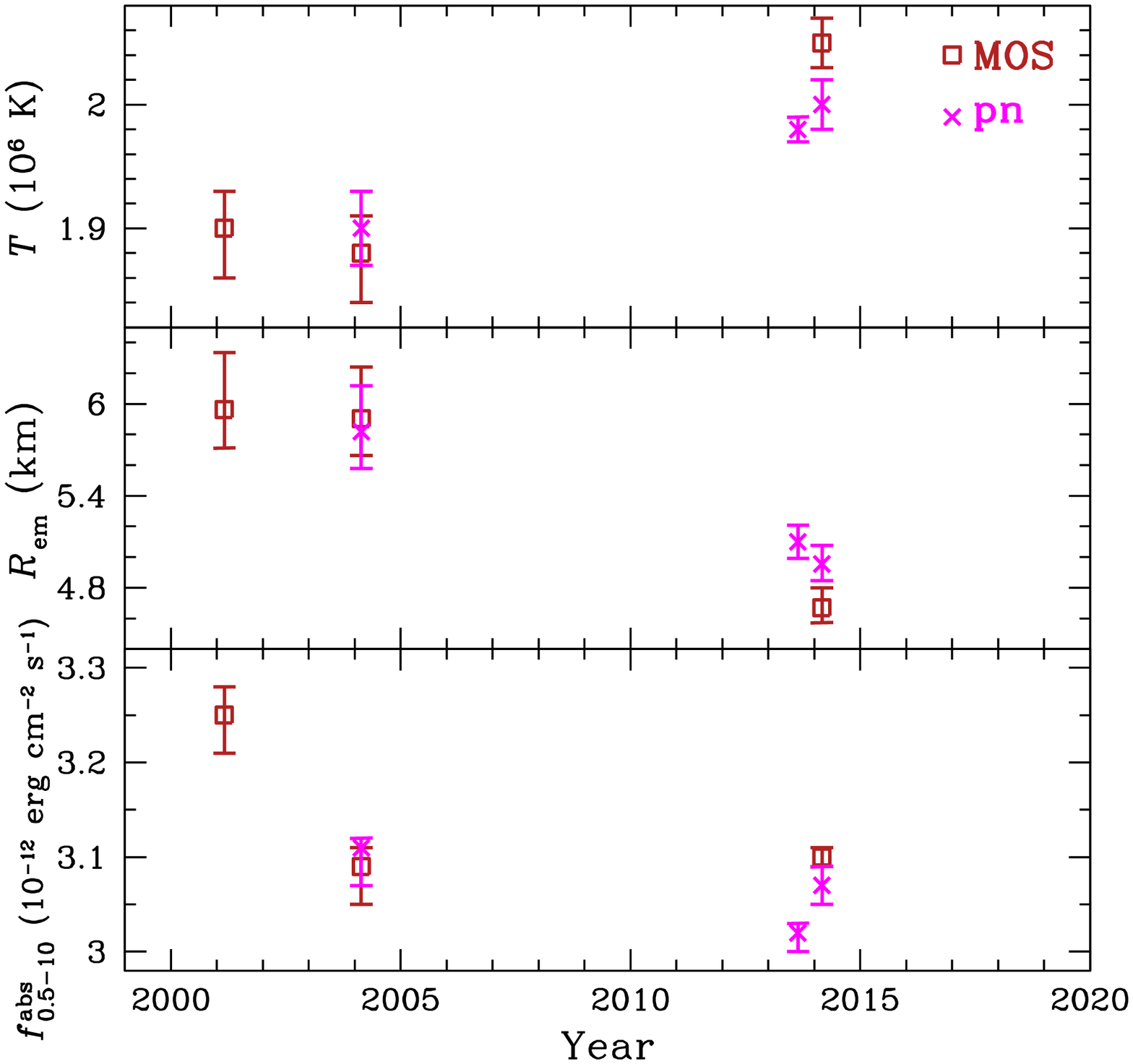}
\includegraphics[width=1.01\columnwidth,trim=0 0 0 20,clip]{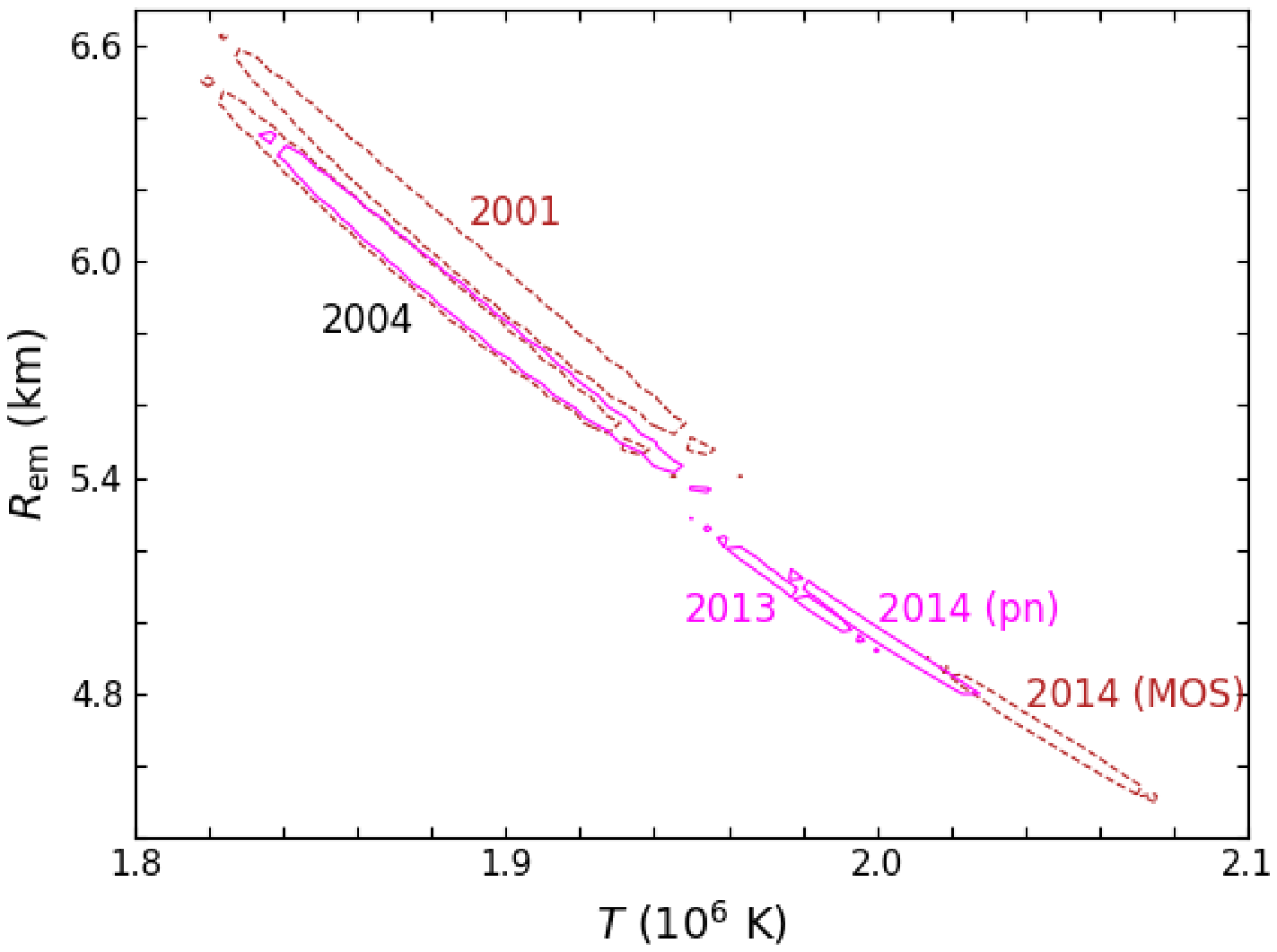}
\caption{
Top: Surface temperature $T$ (upper), emission radius $\Rem$ (middle),
and absorbed 0.5--10~keV flux $\fabs$ (lower) of \wgaj\
from non-magnetic carbon atmosphere model fits to \XMM\ MOS (squares)
and pn (crosses) spectra (see Tables~\ref{tab:xmmmos} and \ref{tab:xmmpn}).
Errors are 1$\sigma$.
Bottom: 90~percent confidence contours for $T$ and $\Rem$ as measured
using MOS (dashed) and pn (solid).  Labels indicate observation year.
}
\label{fig:wgaj}
\end{center}
\vspace{-2em}
\end{figure}

Figure~\ref{fig:spectra_wgaj} shows spectra of \wgaj\ from
\Chandra\ in 2000 and 2014, EPIC-MOS in 2001, 2004, and 2014,
and EPIC-pn in 2004, 2013, and 2014.
As noted in Section~\ref{sec:data}, the 2000 \Chandra\ observation
of the field of \wgaj\ has the CCO located
far off-axis (CCO on S1 chip is 23\arcmin\ from ACIS-I aimpoint;
see also \citealt{lazendicetal03}) and, as a result, the source's
point-spread-function is significantly extended and distorted
(hence the $1\arcmin\times2\arcmin$ diameter source extraction region;
see Section~\ref{sec:data}).
The CCO in the 2014 \Chandra\ observation is also off-axis and distorted.
These off-axis data are also not as well-calibrated.
Therefore to minimize introducing additional systematic effects, we base our discussions and conclusions on results of fits
to the \XMM\ spectra of \wgaj.
Nevertheless we present fits of the \Chandra\ spectra in Appendix~\ref{sec:wgajoff} for the sake of
completeness and to compare with results from previous works.
Furthermore, here we consider \XMM\ MOS and pn data separately;
a joint analysis is presented in Appendix~\ref{sec:wgajoff}.

Tables~\ref{tab:xmmmos} and \ref{tab:xmmpn} show results of model fits
of \XMM\ MOS spectra at 0.2--7~keV and pn spectra at 0.3--10~keV,
respectively.
Fitting spectra with a single absorbed blackbody yields poor results.
When all absorbed thermal model parameters ($\NH$, $T$, and $\Rem$) are
allowed to vary between all observations, the fit is still not good.
Note that the blackbody fit parameters, e.g.,
$\Rem\sim0.5\mbox{ km}$, are comparable to those obtained in a
fit of only the 2001 \XMM\ spectra \citep{cassamchenaietal04}.
On the other hand, good fits can be obtained with atmosphere
model spectra such as those given in Table~\ref{tab:atm}.
Comparisons of fits where all parameters ($\NH$, $T$, and $\Rem$)
are linked or allowed to vary between observations indicate
fits which allow for varying $T$ and $\Rem$ are justified,
e.g., for non-magnetic carbon model fits to
pn spectra, we find F-test probabilities of
$1\times10^{-5}$ for linked parameters versus varying $T$
and $4\times10^{-3}$ for varying $T$ versus varying $T$ and $\Rem$,
while for MOS spectra the F-test probabilities are
$1\times10^{-14}$ for linked parameters versus varying $\NH$
and $4\times10^{-4}$ for varying $\NH$ versus varying $T$ and $\Rem$.
All atmosphere model fits give $\NH\approx(6-7)\times10^{21}\mbox{ cm$^{-2}$}$,
comparable to the SNR absorption.
The best fit to both MOS and pn spectra are non-magnetic carbon
atmospheres, but non-magnetic hydrogen are nearly as good
($\Delta\chi^2=24$ for 978 dof and $\Delta\chi^2=18$ for 1210 dof
for MOS and pn, respectively).
We note that \citet{potekhinetal20} use \texttt{nsx} to fit the
2014 pn spectrum and obtain similar results to our \texttt{nsx}
fit results of the same spectrum (see Table~\ref{tab:xmmpn}).
On the other hand,
while low and high magnetic field hydrogen models yield similar quality
MOS spectral fits to non-magnetic models
(see examples in Table~\ref{tab:xmmmos}), these same magnetic
models do not fit the pn spectra nearly as well.
This can be attributed to the fact that magnetic models are softer
than non-magnetic models for the same effective temperature
\citep{shibanovetal92} and the pn spectra here extends to 10~keV
while MOS spectra only go up to 7~keV.

In Figure~\ref{fig:wgaj}, we show the evolution of the MOS and pn
inferred temperature $T$, emission radius $\Rem$, and absorbed flux
$\fabs$ for non-magnetic carbon atmosphere model fits
(hydrogen atmosphere results show similar trends).
Inferred values at the same or nearly the same epoch are consistent
within $\sim 2\sigma$.
A temperature increase of $\sim4\pm2$~percent and emission radius decrease
of $\sim 14\pm6$~percent are seen to occur between 2004 and 2013--14,
while flux remains approximately constant.
Also evident is an apparent flux drop, but without a significant
change of temperature
and emission radius, between 2001 and 2004.

\subsection{\xmmj} \label{sec:xmmj}

\begin{figure}
\begin{center}
\includegraphics[width=0.95\columnwidth]{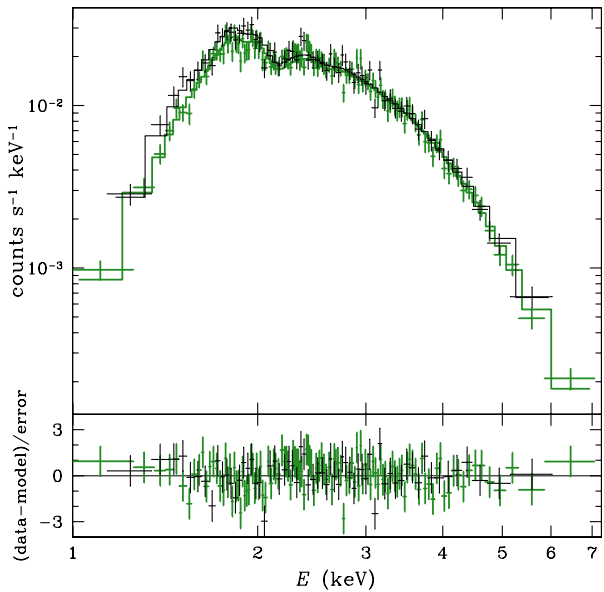}
\caption{
Upper panel shows \Chandra\ spectra (black for 2009 and green for merged 2018
data) of \xmmj\ and absorbed non-magnetic carbon atmosphere model spectra.
Model results shown here are from fits with all parameters linked between
the two sets of observations.  Lower panel shows fit residuals.
}
\label{fig:spectra_xmmj}
\end{center}
\vspace{-2em}
\end{figure}

\begin{table*}
\centering
\caption{Results of fits to the 2009 and (merged) 2018 \Chandra\ spectra of
\xmmj\ at 0.9--7~keV.
Along with \texttt{pileup} and \texttt{tbabs}, fits are performed using
model \texttt{bbodyrad}, \texttt{nsatmos}, \texttt{nsx}, or \texttt{nsmaxg}
with model parameters linked between 2009 and 2018 fits, unless given otherwise.
Other model parameters are fixed at $M=1.4\,M_\odot$, $R=12\mbox{ km}$,
and $d=4.5\mbox{ kpc}$.
Errors are 1$\sigma$.}
\label{tab:spectra_xmmj}
\begin{tabular}{ccccccc}
\hline
Year & $\alpha$ & $\NH$ & $T$ & $\Rem/R$ & $\fabs$ & $\chi^2$/dof \\
 & & ($10^{22}\mbox{ cm$^{-2}$}$) & ($10^6\mbox{ K}$) & or $\mbox{4.5 kpc}/d$ & ($10^{-13}\mbox{ erg cm$^{-2}$ s$^{-1}$}$) & \\
\hline
\multicolumn{7}{c}{blackbody} \\
2009 & $0.42^{+0.10}_{-0.09}$ & $4.7^{+0.1}_{-0.1}$ & $5.85^{+0.09}_{-0.09}$ & $0.066^{+0.003}_{-0.001}$ & $5.6^{+0.2}_{-0.2}$ & 150.4/178 \\
2018 & $0.34^{+0.06}_{-0.06}$ & ---                 & ---                    & ---                       & $5.4^{+0.1}_{-0.1}$ &           \\
\\
\multicolumn{7}{c}{\texttt{nsatmos} - non-magnetic hydrogen} \\
2009 & $0.19^{+0.10}_{-0.10}$ & $5.4^{+0.1}_{-0.1}$ & $3.45^{+0.09}_{-0.08}$ & $0.29^{+0.02}_{-0.02}$ & $5.4^{+0.2}_{-0.3}$ & 154.3/178 \\
2018 & $0.12^{+0.06}_{-0.06}$ & ---                 & ---                    & ---                    & $5.3^{+0.1}_{-0.2}$ &           \\
\\
\multicolumn{7}{c}{\texttt{nsx} - non-magnetic carbon} \\
2009 & $0.25^{+0.09}_{-0.10}$ & $5.6^{+0.1}_{-0.1}$ & $2.33^{+0.08}_{-0.08}$ & $0.77^{+0.10}_{-0.08}$ & $5.4^{+0.2}_{-0.3}$ & 154.2/178 \\
2018 & $0.18^{+0.06}_{-0.06}$ & ---                 & ---                    & ---                    & $5.3^{+0.1}_{-0.2}$ &           \\
\\
\multicolumn{7}{c}{\texttt{nsmaxg} - hydrogen at $10^{13}\mbox{ G}$} \\
2009 & $0.41^{+0.09}_{-0.09}$ & $5.3^{+0.1}_{-0.1}$ & $3.77^{+0.11}_{-0.10}$ & $0.24^{+0.02}_{-0.02}$ & $5.6^{+0.1}_{-0.3}$ & 150.4/178 \\
2018 & $0.33^{+0.05}_{-0.06}$ & ---                 & ---                    & ---                    & $5.4^{+0.1}_{-0.2}$ &           \\
\hline
\end{tabular}
\vspace{-1em}
\end{table*}

Figure~\ref{fig:spectra_xmmj} shows the 2009 and (merged) 2018 \Chandra\
spectra of \xmmj, and
Table~\ref{tab:spectra_xmmj} shows results of model fits of these spectra
at 0.9--7~keV.
For the blackbody model, the fit yields a small hot spot
($\Rem\approx0.8\mbox{ km}$ for $d=4.5\mbox{ kpc}$)
that would suggest pulsations which are not detected despite searches
(pulsed fraction limit of $\sim 30$~percent for spin periods comparable
to other CCOs; see Table~\ref{tab:cco}).
These blackbody results are in approximate agreement with those obtained
in previous studies of the 2009 \Chandra\ spectrum \citep{lovchinskyetal11}
and a 2007 \XMM\ spectrum \citep{gaensleretal08}.

Comparable fits are achieved when an atmosphere model spectrum
is used instead of a blackbody spectrum.
The best-fit model is a magnetic ($B=10^{13}\mbox{ G}$) partially
ionized hydrogen atmosphere, with emission size $\Rem\approx2.9\mbox{ km}$
(for $R=12\mbox{ km}$),
although other magnetic hydrogen models at $B\ge 10^{12}\mbox{ G}$
or a non-magnetic hydrogen model with $\Rem\approx3.5\mbox{ km}$
yield only slightly poorer fits ($\Delta\chi^2\lesssim 4$)
and are good fits to the observed spectra.
A carbon atmosphere with $\Rem\approx9.2\mbox{ km}$ is a slightly
worse fit compared to the best-fit model ($\Delta\chi^2\approx4$).
We also note that the inferred absorption
$\NH\approx(5.3-5.6)\times10^{22}\mbox{ cm$^{-2}$}$
is somewhat greater than
$\NH\approx(3-4)\times10^{22}\mbox{ cm$^{-2}$}$
from spectral analysis of the SNR using \Chandra\ \citep{gaensleretal08},
\Suzaku\ \citep{yasumietal14}, and \XMM\ \citep{lovchinskyetal11}.

The spectral fit results shown in Table~\ref{tab:spectra_xmmj}
assume the nominal distance to \xmmj\ of 4.5~kpc, which is
determined from what appears to be interaction between a molecular
cloud and SNR G350.1$-$0.3 \citep{gaensleretal08}.
On the other hand, \ion{H}{i} absorption towards the SNR gives
a distance range of 4.5--10.7~kpc \citep{gaensleretal08}.
Taking into account the considerations described in Section~\ref{sec:cxoj},
we determine that the distance to \xmmj\ is $5.9\pm0.7\mbox{ kpc}$
if the CCO has a carbon atmosphere with emission size $\Rem=R$,
while a hydrogen atmosphere is possible if the hot region covers
$\sim$40~percent of the neutron star surface at a maximum distance
to the CCO of 10.7~kpc.
We note that \citet{potekhinetal20} fit the early epoch \Chandra\
spectrum with a carbon atmosphere model via \texttt{nsx} and obtain
a comparable distance of $6.1^{+2.6}_{-1.9}\mbox{ kpc}$.

Finally we examine whether there is evolution of model parameters
from fits of the 2009 and 2018 spectra.
We find there is
very little change in the fit statistics (i.e., $\Delta\chi^2<2$;
see, e.g., Section~\ref{sec:cco})
when allowing $\NH$, $T$, and/or $\Rem$ to vary between these data,
and thus the spectral model parameters do not appear to vary
significantly between observations.

\subsection{Cassiopeia~A} \label{sec:casa}

\begin{figure}
\begin{center}
\includegraphics[width=0.95\columnwidth]{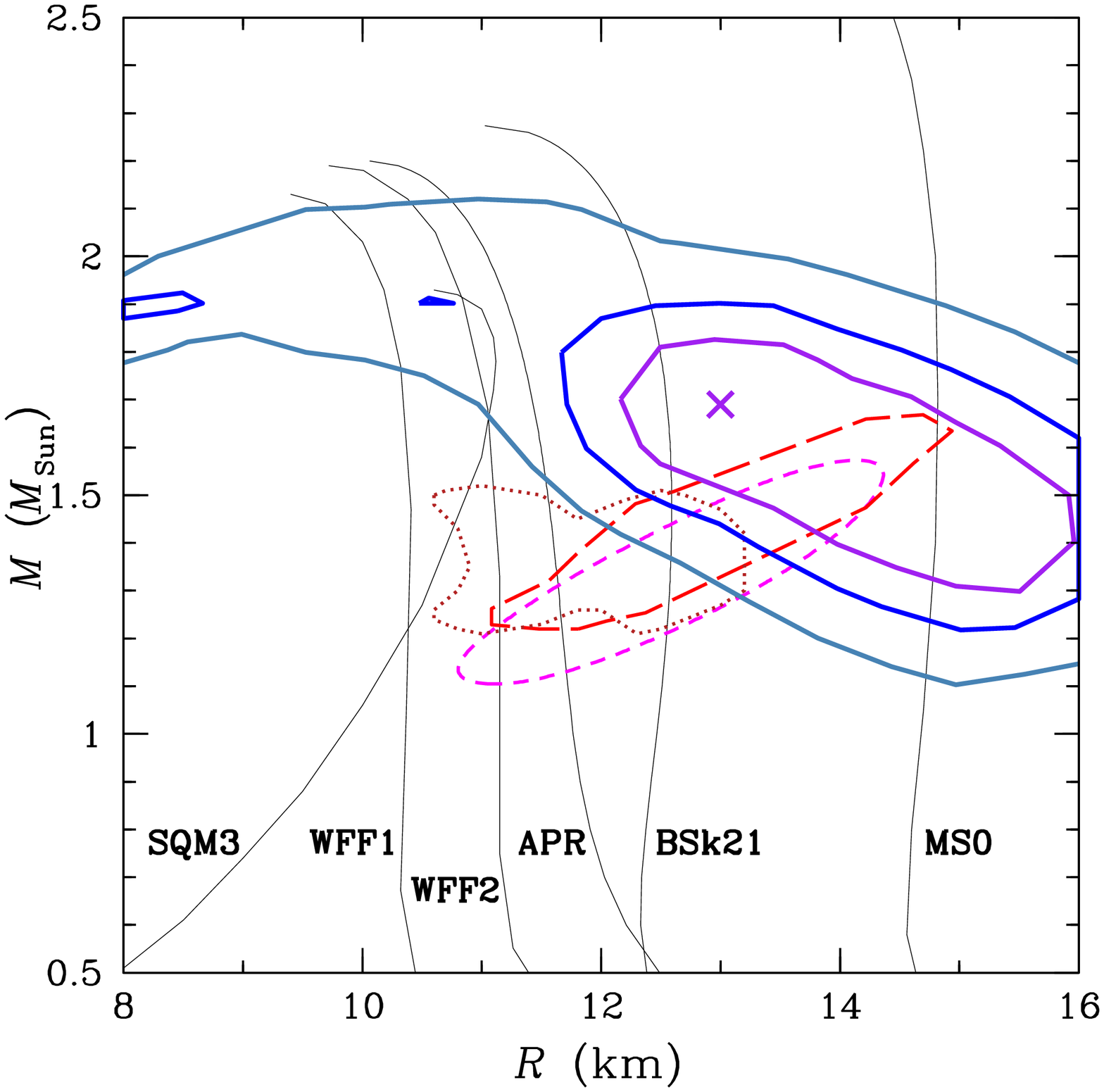}
\caption{
Neutron star mass $M$ and radius $R$.
Solid lines are 1$\sigma$, 90~percent, and 99~percent confidence contours
obtained from fitting \Chandra\ ACIS-S Graded spectra of the
neutron star in the Cassiopeia~A supernova remnant, assuming $\NH$ does
not vary between observations.
Cross marks the best-fitting $M=1.69\,\Msun$ and $R=13.0\mbox{ km}$.
Long-dashed, short-dashed, and dotted lines indicate $M$--$R$ constraints
at 1$\sigma$ of millisecond pulsar PSR~J0030+0451 from \citet{milleretal19} and
\citet{rileyetal19} and of GW170817 from \citet{abbottetal18}, respectively.
The light solid lines indicate the $M$--$R$ relation obtained from
a sample of theoretical nuclear equations of state, in particular,
APR \citep{akmaletal98}, BSk21 \citep{potekhinetal13}, and
MS0, SQM3, WFF1, and WFF2
(see \citealt{lattimerprakash01}, and references therein).
\label{fig:casa_mr}
}
\end{center}
\vspace{-1em}
\end{figure}

To account for a correlation between $M$, $R$, and \texttt{pileup}
grade migration parameter $\alpha$ (see \citealt{shterninetal21}),
we iterate initially between two model fits to the \Chandra\ ACIS-S
Graded spectra of the CCO in Cassiopeia~A.
The first fit allows for varying $M$ and $R$
(with the same values for all 14 observations) and fixed $\alpha_1$
(for observations with frame time of 3.24~s; see Table~\ref{tab:data})
and $\alpha_2$ (for observations with frame time of 3.04~s).
The second fit allows for varying $\alpha_1$ and $\alpha_2$
and fixed $M$ and $R$.
Both fits allow for varying $\NH$
(but tied across and hence the same value for all observations)
and $T$ (can be different for each observation).
The best-fit converges to $M=1.69\,\Msun$, $R=13.0\mbox{ km}$,
and $\NH=1.67^{+0.02}_{-0.06}\times 10^{22}\mbox{ cm$^{-2}$}$,
which are nearly the same as values found in \citet{wijngaardenetal19},
and $\alpha_1=0.33\pm0.03$ and $\alpha_2=0.24\pm0.02$,
which are consistent with values found in
\citet{heinkeho10,hoetal15,shterninetal21}.
Figure~\ref{fig:casa_mr} shows the $M$-$R$ confidence contours.

\begin{figure}
\begin{center}
\includegraphics[width=0.48\textwidth]{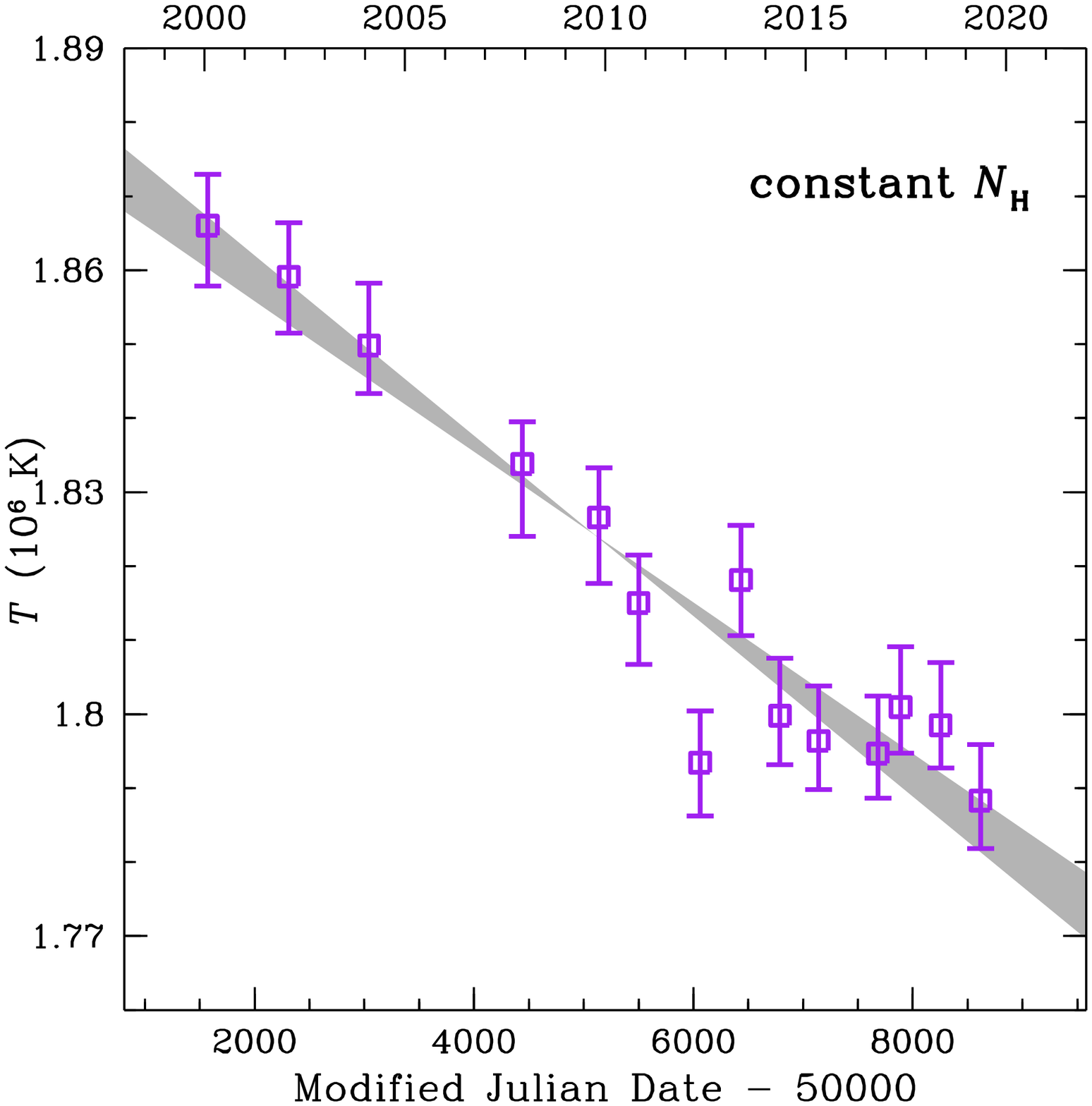}
\includegraphics[width=0.48\textwidth]{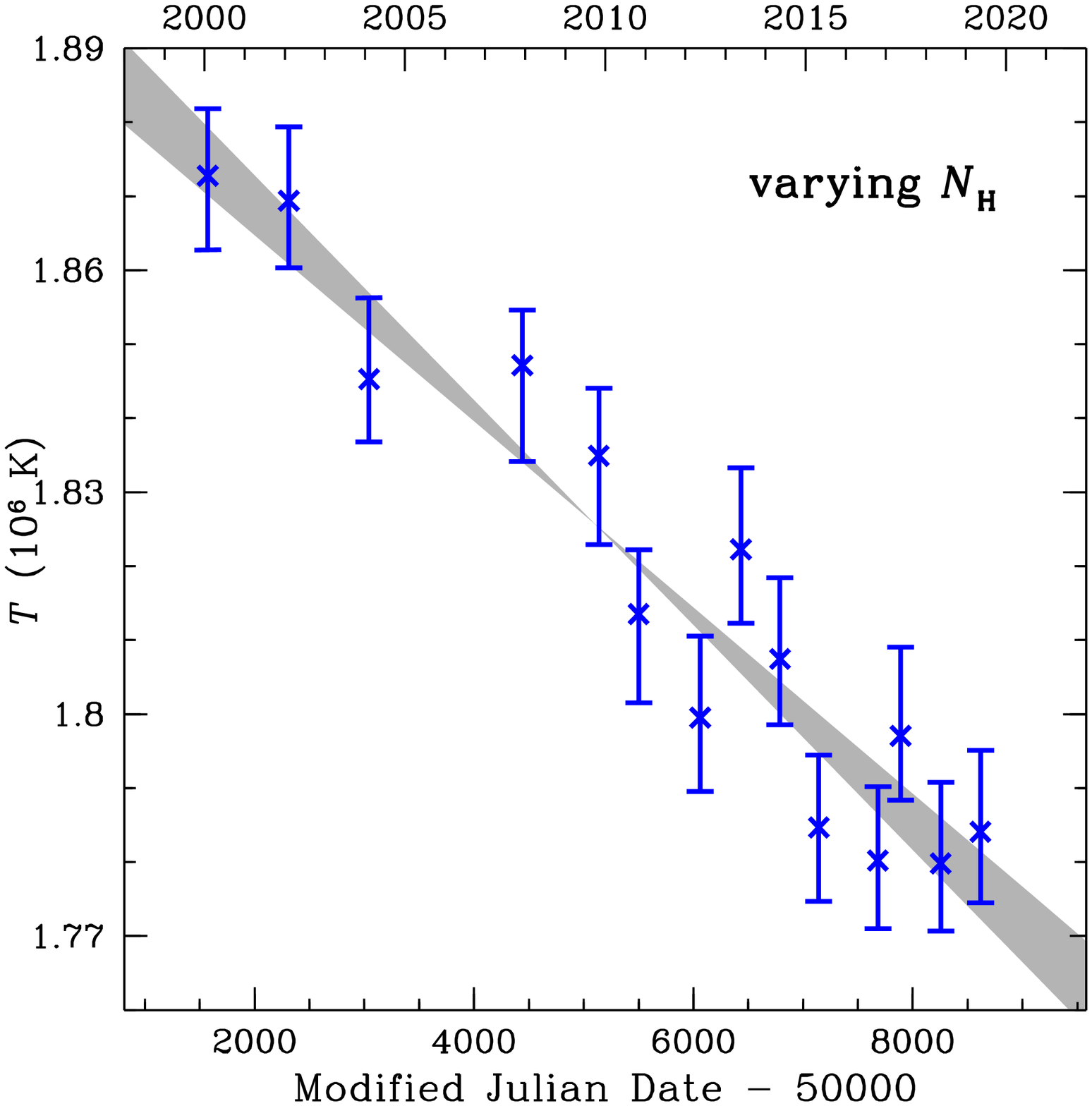}
\caption{
Surface temperature $T$ of the neutron star in the Cassiopeia~A
supernova remnant as measured from \Chandra\ ACIS-S Graded spectra
over the past two decades.
Data points indicate $T$ obtained using the best-fitting neutron star
mass $M=1.69\,\Msun$ and radius $R=13\mbox{ km}$ and a constant
absorption column $\NH=1.67\times 10^{22}\mbox{ cm$^{-2}$}$ (top panel)
and varying $\NH$ (bottom panel).
Error bars are 1$\sigma$.
Shaded regions show the uncertainty range for a linear fit to each
set of $T$, i.e, a ten-year decline rate of $2.2\pm0.2$~percent
for constant $\NH$ and $2.8\pm0.3$~percent for varying $\NH$,
with the fit centered at the mid-point of all observations (MJD~55095).
\label{fig:casa_temp}
}
\end{center}
\vspace{-1em}
\end{figure}

\begin{figure}
\begin{center}
\includegraphics[width=1.05\columnwidth]{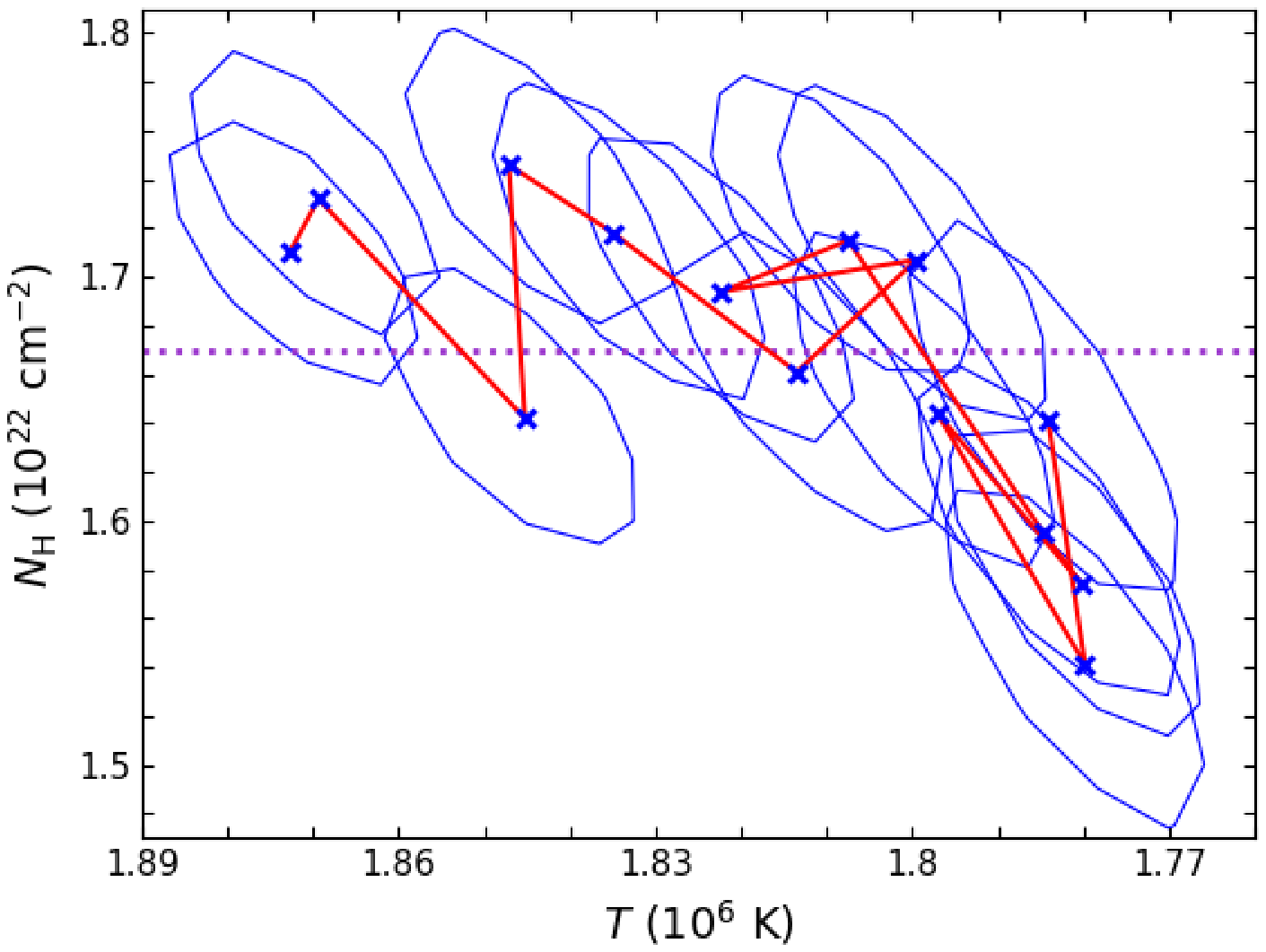}
\caption{
Absorption column $\NH$ and surface temperature $T$ as measured
from a fit to \Chandra\ ACIS-S Graded spectra of the Cassiopeia~A
neutron star.
Crosses and 1$\sigma$ confidence contours indicate $\NH$--$T$ fit values
using the best-fitting neutron star mass $M=1.69\,\Msun$ and radius
$R=13\mbox{ km}$, with lines connecting consecutive observations,
while the horizontal dotted line indicates constant
$\NH=1.67\times 10^{22}\mbox{ cm$^{-2}$}$
used in the other spectral fit (see Figure~\ref{fig:casa_temp}).
\label{fig:casa_nh}
}
\end{center}
\vspace{-1em}
\end{figure}

Our final spectral fits allow for a varying surface temperature
$T$ between each observation and either assumes a constant
absorption $\NH=1.67\times 10^{22}\mbox{ cm$^{-2}$}$ or allows
$\NH$ to vary between each observation, while holding $M$,
$R$, $\alpha_1$, and $\alpha_2$ at the best-fit values from above.
Results are given in Table~\ref{tab:casa} and Figures~\ref{fig:casa_temp}
and \ref{fig:casa_nh}.
A F-fest between the fits with a constant and varying $\NH$ yields
a probability of 0.02, so allowing $\NH$ to vary is possibly preferred.
The ten-year cooling rates of $2.2\pm0.2$~percent for constant $\NH$ and
$2.8\pm0.3$~percent for varying $\NH$ are consistent with those from
\citet{wijngaardenetal19} and cooling rate upper limits of 2.4~percent
for constant $\NH$ and 3.3~percent for varying $\NH$ from
\citet{posseltpavlov18}.

\begin{table*}
\caption{
Results of fits to \Chandra\ ACIS-S Graded spectra of the CCO in Cassiopeia~A.
Two spectral fit results using \texttt{nsx} are shown,
one with constant $\NH$ and one with varying $\NH$.
Other model parameters are fixed at $M=1.69\,M_\odot$, $R=13\mbox{ km}$,
$d=3.4\mbox{ kpc}$,
and $\alpha=0.33$ or $0.24$ for observations with a frame time of 3.24~s
(i.e., ObsIDs 114, 1952, and 5196; see Table~\ref{tab:data}) or 3.04~s,
respectively.
Each set of 14 temperatures and absorbed 0.5-7~keV fluxes $f^{\rm abs}_{0.5-7}$
are fit to a linear decline, with decline rate and fit statistic as shown.
For merged ObsIDs, the MJD listed is that of the first ObsID.
Error bars are 1$\sigma$.
}
\label{tab:casa}
\begin{tabular}{rlcccccc}
\hline
 & & $\NH$ & \multicolumn{2}{c}{$1.67\times10^{22}\mbox{ cm$^{-2}$}$} & \multicolumn{3}{c}{see below} \\
 & & $\chi^2$/dof & \multicolumn{2}{c}{1654/1641} & \multicolumn{3}{c}{1627/1627} \\ \\
ObsID & Date & MJD & $T$ & $f_{\rm0.5-7}^{\rm abs}$ & $T$ & $f_{\rm0.5-7}^{\rm abs}$ & $\NH$ \\
 & & & ($10^6\mbox{ K}$) & ($10^{-13}\mbox{ erg cm$^{-2}$ s$^{-1}$}$) & ($10^6\mbox{ K}$) & ($10^{-13}\mbox{ erg cm$^{-2}$ s$^{-1}$}$) & ($10^{22}\mbox{ cm$^{-2}$}$) \\
\hline
114 & 2000 Jan 30 & 51573.4 & 1.866$^{+0.007}_{-0.008}$ & 7.5$^{+0.2}_{-0.1}$ & 1.873$^{+0.009}_{-0.010}$ & 7.5$^{+0.2}_{-0.1}$ & 1.71$^{+0.03}_{-0.04}$ \\
1952 & 2002 Feb 6 & 52311.3 & 1.859$^{+0.007}_{-0.008}$ & 7.4$^{+0.2}_{-0.2}$ & 1.869$^{+0.010}_{-0.009}$ & 7.4$^{+0.1}_{-0.1}$ & 1.73$^{+0.04}_{-0.03}$ \\
5196 & 2004 Feb 8 & 53043.7 & 1.850$^{+0.008}_{-0.006}$ & 7.2$^{+0.1}_{-0.1}$ & 1.845$^{+0.011}_{-0.008}$ & 7.2$^{+0.2}_{-0.1}$ & 1.64$^{+0.04}_{-0.03}$ \\
9117/9773 & 2007 Feb 5/8 & 54439.9 & 1.834$^{+0.006}_{-0.010}$ & 6.7$^{+0.1}_{-0.1}$ & 1.847$^{+0.007}_{-0.013}$ & 6.8$^{+0.2}_{-0.1}$ & 1.75$^{+0.03}_{-0.05}$ \\
10935/12020 & 2009 Nov 2/3 & 55137.9 & 1.827$^{+0.007}_{-0.009}$ & 6.6$^{+0.1}_{-0.1}$ & 1.835$^{+0.009}_{-0.012}$ & 6.7$^{+0.1}_{-0.2}$ & 1.72$^{+0.04}_{-0.04}$ \\
10936/13177 & 2010 Oct 31/Nov 2 & 55500.2 & 1.815$^{+0.006}_{-0.008}$ & 6.4$^{+0.1}_{-0.1}$ & 1.813$^{+0.009}_{-0.012}$ & 6.4$^{+0.2}_{-0.1}$ & 1.66$^{+0.04}_{-0.04}$ \\
14229 & 2012 May 15 & 56062.4 & 1.793$^{+0.007}_{-0.007}$ & 6.2$^{+0.1}_{-0.1}$ & 1.800$^{+0.011}_{-0.010}$ & 6.3$^{+0.1}_{-0.1}$ & 1.71$^{+0.05}_{-0.04}$ \\
14480 & 2013 May 20 & 56432.6 & 1.818$^{+0.007}_{-0.007}$ & 6.5$^{+0.1}_{-0.1}$ & 1.822$^{+0.011}_{-0.010}$ & 6.6$^{+0.2}_{-0.1}$ & 1.69$^{+0.04}_{-0.04}$ \\
14481 & 2014 May 12 & 56789.1 & 1.800$^{+0.008}_{-0.007}$ & 6.2$^{+0.2}_{-0.1}$ & 1.807$^{+0.011}_{-0.009}$ & 6.3$^{+0.1}_{-0.1}$ & 1.71$^{+0.04}_{-0.04}$ \\
14482 & 2015 Apr 30 & 57142.5 & 1.796$^{+0.007}_{-0.007}$ & 6.2$^{+0.1}_{-0.1}$ & 1.785$^{+0.010}_{-0.010}$ & 6.2$^{+0.1}_{-0.2}$ & 1.59$^{+0.04}_{-0.04}$ \\
19903/18344 & 2016 Oct 20/21 & 57681.2 & 1.795$^{+0.008}_{-0.006}$ & 6.2$^{+0.1}_{-0.1}$ & 1.780$^{+0.010}_{-0.009}$ & 6.1$^{+0.1}_{-0.1}$ & 1.57$^{+0.04}_{-0.04}$ \\
19604 & 2017 May 16 & 57889.7 & 1.801$^{+0.008}_{-0.006}$ & 6.4$^{+0.1}_{-0.1}$ & 1.797$^{+0.012}_{-0.009}$ & 6.3$^{+0.1}_{-0.1}$ & 1.64$^{+0.05}_{-0.04}$ \\
19605 & 2018 May 15 & 58253.7 & 1.798$^{+0.008}_{-0.006}$ & 6.3$^{+0.1}_{-0.2}$ & 1.780$^{+0.011}_{-0.009}$ & 6.3$^{+0.1}_{-0.2}$ & 1.54$^{+0.05}_{-0.04}$ \\
19606 & 2019 May 13 & 58616.5 & 1.788$^{+0.008}_{-0.006}$ & 6.2$^{+0.2}_{-0.1}$ & 1.784$^{+0.011}_{-0.010}$ & 6.2$^{+0.1}_{-0.1}$ & 1.64$^{+0.05}_{-0.05}$ \\
\hline
 & \multicolumn{2}{r}{10-year decline rate} & $2.2\pm0.2\%$ & $10\pm1\%$ & $2.8\pm0.3\%$ & $11\pm1\%$ & \\
 & & $\chi^2$/dof & 13.4/12 & 14.7/12 & 10.0/12 & 12.2/12 & \\
\hline
\end{tabular}
\vspace{-1em}
\end{table*}

\vspace{-1em}
\section{Discussion} \label{sec:discuss}

In this work, we analyze \Chandra\ and \XMM\ CCD spectra taken over one or
two decades of four of the youngest CCOs known,
\cxoj\ in G330.2+1.0 with an age of $<1000\mbox{ yr}$,
\wgaj\ in G347.3$-$0.5 with an age of 1500--2300~yr,
\xmmj\ in G350.1$-$0.3 with an age of $<600\mbox{ yr}$, and
Cassiopeia~A with an age of 340~yr.
Spectra of each CCO can be well-described by neutron star atmosphere models.
For \cxoj, the spectral fits do not allow us to distinguish between
a non-magnetic carbon atmosphere or a hydrogen atmosphere at magnetic
fields $B\le7\times10^{10}\mbox{ G}$ or $B\ge10^{12}\mbox{ G}$.
For \wgaj, the best-fit spectral model is a non-magnetic atmosphere
composed of carbon or possibly hydrogen.
For \xmmj, the best-fit model is a magnetic ($B\ge10^{13}~\mbox{ G}$)
hydrogen atmosphere, although non-magnetic hydrogen or carbon atmosphere
spectra yield nearly as good fits to observed spectra.
For Cassiopeia~A, a non-magnetic carbon atmosphere spectrum continues
to be a good fit to the spectra of the CCO.

Based purely on quality of fits of atmosphere model spectra to observed
spectra obtained here, one composition is not preferred over another
for the three older CCOs.
The primary differences in fit results when the atmosphere is
composed of carbon instead of hydrogen are a $\sim 40$~percent
lower temperature and $\sim 3$ times larger emission radius
(to maintain nearly constant $\Rem^2T^4$).
The larger $\Rem$ could be an argument in favor of carbon because
it is closer to the neutron star radius $R$, which would imply
that the entire surface is essentially at a single temperature and
would explain non-detection thus far of pulsations from each of
these three CCOs.
However, a hot region with radius 3--4 times smaller than $R$ can still
produce a pulsed fraction below current limits of 20--40~percent
in the spin period range $0.1-0.4\mbox{ s}$ of known CCOs
(see Section~\ref{sec:intro};
for pulsed fraction dependence on spot size, see, e.g.,
\citealt{psaltisetal00,dedeoetal01,weinbergetal01,bogdanovetal08,lambetal09,baubocketal15,elshamoutyetal16}).
For example, \citet{gotthelfetal10} find that a model which includes
a hot spot with $\Rem/R\sim0.4$ (and a second smaller spot) is able
to produce a pulsed fraction
that matches the 11~percent measured for the CCO in Puppis~A.
Also of note is the stronger limit on the Cassiopeia~A CCO pulsed
fraction of 12~percent for $P>0.01\mbox{ s}$ compared to the other CCOs
(see Table~\ref{tab:cco}).
Meanwhile,
a hydrogen atmosphere is a natural consequence of even a very low-level
of accretion from the interstellar medium on to a relatively cool
neutron star surface several hundred years after neutron star formation
(see Section~\ref{sec:cco} for further discussion).
Thus from an evolution standpoint, a hydrogen atmosphere for the older
CCOs studied here might be preferred.
Future measurements of pulsations or improvements to pulsation
constraints could provide stronger indications of their atmosphere
composition.

\vspace{-1em}
\subsection{Early evolution of CCOs} \label{sec:cco}

Hereafter we assume the four CCOs studied here are born similar
to one another, even though their intrinsic properties and evolution
must vary somewhat.
This assumption enables us to use each as a snapshot of what a
prototypical CCO might look like at different stages in its evolution.
The ages derived for SNRs G330.2+1.0 and G350.1$-$0.3 are based on
expansion of ejecta/shell material at constant velocity, and
since deceleration is likely to occur, their ages are upper limits
\citep{borkowskietal18,borkowskietal20}.
Nevertheless for simplicity, we will consider these ages to be accurate,
such that
Cassiopeia~A is the youngest CCO of the four at 340~yr old,
\xmmj\ is the next youngest at 600~yr old, and \cxoj\ is 1000~yr old.
\wgaj\ is the oldest of the four, with an age of 1500--2300~yr.
Note that the other three CCOs highlighted in Section~\ref{sec:intro}
are much older, with ages $>4000$~yr
\citep{rogeretal88,sunetal04,mayeretal20}.

While the youngest CCO, Cassiopeia~A, likely has a carbon atmosphere,
the three older CCOs studied in the present work might have a
hydrogen atmosphere,
especially given the relatively weak pulsation constraint for each
of these three.
This evolution in atmosphere composition is due initially to the hot
neutron star at birth.
The high temperatures prompt formation of carbon and other heavy elements
from nuclear fusion of any residual surface hydrogen and helium
\citep{changbildsten03,changbildsten04,changetal10,wijngaardenetal19,wijngaardenetal20}.
These nuclear reactions cease to be effective after several hundred years
as the neutron star surface cools.
Then even a small amount of accretion from the interstellar medium
leads to sufficient hydrogen to form an optically thick atmosphere
\citep{blaesetal92,wijngaardenetal19}.

\begin{figure}
\begin{center}
\includegraphics[width=0.45\textwidth]{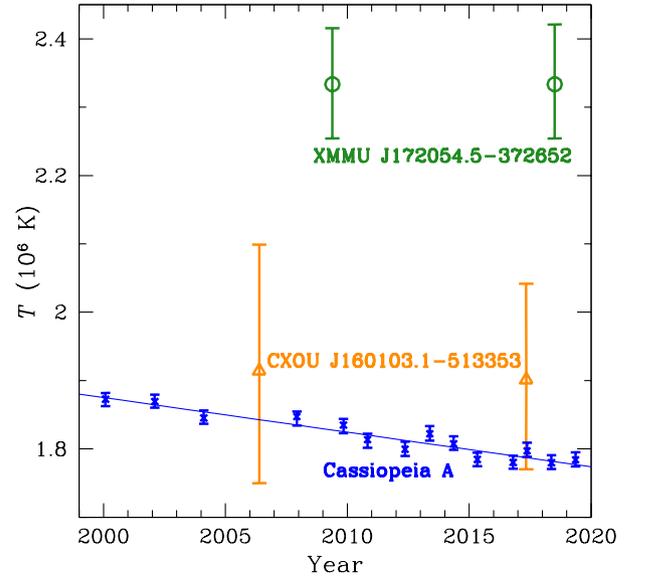}
\caption{
Surface temperature $T$ of \cxoj\ (triangles) and \xmmj\ (circles)
from non-magnetic carbon atmosphere model fits to \Chandra\ spectra
(see Table~\ref{tab:spectra_dt}); see Figure~\ref{fig:wgaj} for \wgaj.
Temperatures of the CCO in Cassiopeia~A (crosses)
are from the carbon atmosphere model fit with varying $\NH$
(see Table~\ref{tab:casa});
line shows the best-fit linear decline rate of 2.8~percent per decade.
Errors are 1$\sigma$.
}
\label{fig:temp}
\end{center}
\vspace{-1em}
\end{figure}

Due to their high core temperatures at birth,
neutron stars cool predominantly by neutrino emission
at ages $\lesssim 10^6\mbox{ yr}$ \citep{potekhinetal15}.
The most rapid changes of surface temperature occur early, first
when the temperature of the outer layers achieves equilibrium with
the rapidly cooling core at an age of $\lesssim 100\mbox{ yr}$
(\citealt{lattimeretal94,gnedinetal01}; see also \citealt{nomototsuruta87})
and then when the temperature drops below the critical temperature
for core neutrons to become superfluid, which activates the efficient
neutrino emission process of Cooper pair formation and breaking
\citep{gusakovetal04,pageetal04}.
The rapid cooling of the CCO in Cassiopeia~A
(at a ten-year rate of $\approx 2.2\pm0.2$ or $2.8\pm0.3$~percent, depending
on whether $\NH$ varies between observations)
indicates the latter starts to take place at an age of $\sim 200\mbox{ yr}$
\citep{pageetal11,shterninetal11}.
Neutron star cooling models predict that by an age of several
hundred years, the cooling rate will be $<1$~percent per decade.
From the 1$\sigma$ temperature uncertainties of our fit results
with model parameters linked between observations
(see Tables~\ref{tab:spectra_cxoj} and \ref{tab:spectra_xmmj}),
we estimate upper limits on the ten-year cooling rates of 6~percent
for \xmmj\ and 17~percent for \cxoj.
We measure a possible increase in temperature of $\sim4\pm2$~percent
(accompanied by a decrease in emission area; see Table~\ref{tab:xmmmos}
and \ref{tab:xmmpn}) for \wgaj.
We also perform fits of the spectra of \xmmj\ and \cxoj\ which allow
the temperature to be different between each observation,
and the results are shown in Table~\ref{tab:spectra_dt} and
Figure~\ref{fig:temp} (analogous results for \wgaj\ are shown in
Figure~\ref{fig:wgaj}).
We point out that we are concerned in Figure~\ref{fig:temp} with relative
changes in temperature and not in absolute temperature differences between
CCOs since absolute temperatures depend on a variety of factors that
are intrinsic to each CCO and may be different among CCOs in our sample,
e.g., mass (and hence neutrino cooling rate) and radius
(and hence gravitational redshift) and envelope composition and thickness.
Nevertheless it is noteworthy that the temperatures of all three
older CCOs appear to be higher than those of Cassiopeia~A.
Unlike for the 340~yr old CCO in Cassiopeia~A, we do not see that
temperatures of the 600~yr old \xmmj\ and 1000~yr old \cxoj\
are changing, at least within measurement uncertainties.

\begin{table}
\centering
\caption{Results of fits to \Chandra\ spectra of \cxoj\ and \xmmj\
using a non-magnetic carbon atmosphere model (\texttt{nsx})
and allowing $T$ to vary between the two epochs of observation,
while $\NH$ and $\Rem$ are linked.
For \xmmj, pile-up parameter $\alpha=0.24^{+0.10}_{-0.10}$
and $0.18^{+0.06}_{-0.06}$ for the 2009 and 2018 fits, respectively.
Other model parameters are fixed at $M=1.4\,M_\odot$, $R=12\mbox{ km}$,
and $d=5\mbox{ kpc}$ for \cxoj\ and $d=4.5\mbox{ kpc}$ for \xmmj.
Absorbed 0.5--10~keV flux $\fabs$ is in $10^{-13}\mbox{ erg cm$^{-2}$ s$^{-1}$}$.
Errors are 1$\sigma$.}
\label{tab:spectra_dt}
\begin{tabular}{cccccc}
\hline
Year & $\NH$ & $T$ & $\Rem/R$ & $\fabs$ & $\chi^2$/dof \\
 & ($10^{22}\mbox{ cm$^{-2}$}$) & ($10^6\mbox{ K}$) & & & \\
\hline
\multicolumn{6}{c}{\cxoj} \\
2006 & $4.8^{+0.3}_{-0.3}$ & $1.9^{+0.2}_{-0.1}$ & $0.75^{+0.25}_{-0.17}$ & $1.5^{+0.1}_{-0.4}$ & 83.5/66 \\
2017 & ---                 & $1.9^{+0.2}_{-0.1}$ & ---                    & $1.4^{+0.1}_{-0.3}$ &         \\
\\
\multicolumn{6}{c}{\xmmj} \\
2009 & $5.6^{+0.1}_{-0.1}$ & $2.34^{+0.08}_{-0.08}$ & $0.77^{+0.10}_{-0.08}$ & $5.4^{+0.2}_{-0.4}$ & 154.1/177 \\
2018 & ---                 & $2.33^{+0.08}_{-0.08}$ & ---                    & $5.3^{+0.1}_{-0.3}$ &         \\
\hline
\end{tabular}
\vspace{-1em}
\end{table}

From our spectral fits, we find a non-magnetic carbon or hydrogen
atmosphere provides a good model for the four youngest known CCOs:
Cassiopeia~A, \xmmj, \cxoj, and \wgaj.
For the oldest of these four, \wgaj,
a relatively low but slowly increasing magnetic
field could explain its (possibly) increasing temperature, as its shrinking
hot spot compresses the magnetic flux and heat from the interior
is conducted along the growing magnetic field.
It is also interesting to note that an outlier signal detected in
gravitational wave searches of \wgaj\ would imply a magnetic field
$\sim 6\times 10^{11}\mbox{ G}$, assuming that the frequency of the
signal is the spin frequency of the CCO \citep{papaetal20}.
One can also consider the three (older) CCOs with a measured spin
period and spin period derivative that indicate a global dipolar
magnetic field strength of $3\times 10^{10}-10^{11}\mbox{ G}$.
Two of these three have hot spot emission that might indicate much
stronger localized fields
\citep{gotthelfetal10,shabaltaslai12,bogdanov14},
while the third undergoes spin frequency glitches, which are
phenomena seen in many young pulsars with stronger fields
\citep{gotthelfhalpern20}.
Magnetic field evolution in the above CCOs can follow from a scenario
where a strong global magnetic field at birth is submerged below the
neutron star surface and gradually diffuses to the surface with an emergence
timescale that depends on burial depth and can be $\sim 1000\mbox{ yr}$
\citep{ho11,ho15,viganopons12,torresforneetal16,fraijaetal18,gourgouliatosetal20}

Previous and future searches for CCOs and their descendants
\citep{gotthelfetal13b,bogdanovetal14,luoetal15,pires18}
will be important for revealing where the CCO class sits within
the general population of neutron stars.
As we show here, studies of CCOs can potentially provide valuable
insights into the surface composition, magnetic field, and
evolution of neutron stars.

\section*{Acknowledgements}

We thank J. Lazendic-Galloway, D. Patnaude, and P. Slane for support
in obtaining \Chandra\ data used here.
WCGH appreciates use of computer facilities at the Kavli Institute for
Particle Astrophysics and Cosmology.
COH acknowledges support from the Natural Sciences and Engineering
Research Council of Canada (NSERC) via Discovery Grant RGPIN-2016-04602.
DLK is supported by the NANOGrav Physics Frontiers Center through NSF
award number 1430284.
This research has made use of data obtained from the Chandra Data Archive
and the Chandra Source Catalog, and software provided by the Chandra
X-ray Center (CXC) in the application packages CIAO, ChIPS, and Sherpa.
This research has made use of data and/or software provided by the High
Energy Astrophysics Science Archive Research Center (HEASARC),
which is a service of the Astrophysics Science Division at NASA/GSFC and
the High Energy Astrophysics Division of the Smithsonian Astrophysical
Observatory.
The authors acknowledge use of the XMM-Newton Science Archive (XSA) for
searching and downloading data.


\section*{Data availability}

The data underlying this article will be shared on reasonable
request to the corresponding author.


\bibliographystyle{mnras}
\bibliography{cco21}

\begin{thebibliography}{}
\makeatletter
\relax
\def\mn@urlcharsother{\let\do\@makeother \do\$\do\&\do\#\do\^\do\_\do\%\do\~}
\def\mn@doi{\begingroup\mn@urlcharsother \@ifnextchar [ {\mn@doi@}
  {\mn@doi@[]}}
\def\mn@doi@[#1]#2{\def\@tempa{#1}\ifx\@tempa\@empty \href
  {http://dx.doi.org/#2} {doi:#2}\else \href {http://dx.doi.org/#2} {#1}\fi
  \endgroup}
\def\mn@eprint#1#2{\mn@eprint@#1:#2::\@nil}
\def\mn@eprint@arXiv#1{\href {http://arxiv.org/abs/#1} {{\tt arXiv:#1}}}
\def\mn@eprint@dblp#1{\href {http://dblp.uni-trier.de/rec/bibtex/#1.xml}
  {dblp:#1}}
\def\mn@eprint@#1:#2:#3:#4\@nil{\def\@tempa {#1}\def\@tempb {#2}\def\@tempc
  {#3}\ifx \@tempc \@empty \let \@tempc \@tempb \let \@tempb \@tempa \fi \ifx
  \@tempb \@empty \def\@tempb {arXiv}\fi \@ifundefined
  {mn@eprint@\@tempb}{\@tempb:\@tempc}{\expandafter \expandafter \csname
  mn@eprint@\@tempb\endcsname \expandafter{\@tempc}}}

\bibitem[\protect\citeauthoryear{{Abbott} et~al.,}{{Abbott}
  et~al.}{2018}]{abbottetal18}
{Abbott} B.~P.,  et~al., 2018, \mn@doi [\prl] {10.1103/PhysRevLett.121.161101},
  \href {https://ui.adsabs.harvard.edu/abs/2018PhRvL.121p1101A} {121, 161101}

\bibitem[\protect\citeauthoryear{{Acero}, {Ballet}, {Decourchelle},
  {Lemoine-Goumard}, {Ortega}, {Giacani}, {Dubner}  \&
  {Cassam-Chena{\"\i}}}{{Acero} et~al.}{2009}]{aceroetal09}
{Acero} F.,  {Ballet} J.,  {Decourchelle} A.,  {Lemoine-Goumard} M.,  {Ortega}
  M.,  {Giacani} E.,  {Dubner} G.,   {Cassam-Chena{\"\i}} G.,  2009, \mn@doi
  [\aap] {10.1051/0004-6361/200811556}, \href
  {https://ui.adsabs.harvard.edu/abs/2009A&A...505..157A} {505, 157}

\bibitem[\protect\citeauthoryear{{Acero}, {Katsuda}, {Ballet}  \&
  {Petre}}{{Acero} et~al.}{2017}]{aceroetal17}
{Acero} F.,  {Katsuda} S.,  {Ballet} J.,   {Petre} R.,  2017, \mn@doi [\aap]
  {10.1051/0004-6361/201629618}, \href
  {https://ui.adsabs.harvard.edu/abs/2017A&A...597A.106A} {597, A106}

\bibitem[\protect\citeauthoryear{{Akmal}, {Pandharipande}  \&
  {Ravenhall}}{{Akmal} et~al.}{1998}]{akmaletal98}
{Akmal} A.,  {Pandharipande} V.~R.,   {Ravenhall} D.~G.,  1998, \mn@doi [\prc]
  {10.1103/PhysRevC.58.1804}, \href
  {https://ui.adsabs.harvard.edu/abs/1998PhRvC..58.1804A} {58, 1804}

\bibitem[\protect\citeauthoryear{{Alcock} \& {Illarionov}}{{Alcock} \&
  {Illarionov}}{1980}]{alcockillarionov80}
{Alcock} C.,  {Illarionov} A.,  1980, \mn@doi [\apj] {10.1086/157656}, \href
  {https://ui.adsabs.harvard.edu/abs/1980ApJ...235..534A} {235, 534}

\bibitem[\protect\citeauthoryear{{Arnaud}}{{Arnaud}}{1996}]{arnaud96}
{Arnaud} K.~A.,  1996, in {Jacoby} G.~H.,  {Barnes} J.,  eds,  Astronomical
  Society of the Pacific Conference Series Vol. 101, Astronomical Data Analysis
  Software and Systems V. p.~17

\bibitem[\protect\citeauthoryear{{Baub{\"o}ck}, {Psaltis}  \&
  {{\"O}zel}}{{Baub{\"o}ck} et~al.}{2015}]{baubocketal15}
{Baub{\"o}ck} M.,  {Psaltis} D.,   {{\"O}zel} F.,  2015, \mn@doi [\apj]
  {10.1088/0004-637X/811/2/144}, \href
  {https://ui.adsabs.harvard.edu/abs/2015ApJ...811..144B} {811, 144}

\bibitem[\protect\citeauthoryear{{Bignami}, {Caraveo}, {De Luca}  \&
  {Mereghetti}}{{Bignami} et~al.}{2003}]{bignamietal03}
{Bignami} G.~F.,  {Caraveo} P.~A.,  {De Luca} A.,   {Mereghetti} S.,  2003,
  \mn@doi [\nat] {10.1038/nature01703}, \href
  {https://ui.adsabs.harvard.edu/abs/2003Natur.423..725B} {423, 725}

\bibitem[\protect\citeauthoryear{{Blaes}, {Blandford}, {Madau}  \&
  {Yan}}{{Blaes} et~al.}{1992}]{blaesetal92}
{Blaes} O.~M.,  {Blandford} R.~D.,  {Madau} P.,   {Yan} L.,  1992, \mn@doi
  [\apj] {10.1086/171955}, \href
  {https://ui.adsabs.harvard.edu/abs/1992ApJ...399..634B} {399, 634}

\bibitem[\protect\citeauthoryear{{Blaschke}, {Grigorian}, {Voskresensky}  \&
  {Weber}}{{Blaschke} et~al.}{2012}]{blaschkeetal12}
{Blaschke} D.,  {Grigorian} H.,  {Voskresensky} D.~N.,   {Weber} F.,  2012,
  \mn@doi [\prc] {10.1103/PhysRevC.85.022802}, \href
  {https://ui.adsabs.harvard.edu/abs/2012PhRvC..85b2802B} {85, 022802}

\bibitem[\protect\citeauthoryear{{Bogdanov}}{{Bogdanov}}{2014}]{bogdanov14}
{Bogdanov} S.,  2014, \mn@doi [\apj] {10.1088/0004-637X/790/2/94}, \href
  {https://ui.adsabs.harvard.edu/abs/2014ApJ...790...94B} {790, 94}

\bibitem[\protect\citeauthoryear{{Bogdanov}, {Grindlay}  \&
  {Rybicki}}{{Bogdanov} et~al.}{2008}]{bogdanovetal08}
{Bogdanov} S.,  {Grindlay} J.~E.,   {Rybicki} G.~B.,  2008, \mn@doi [\apj]
  {10.1086/592341}, \href
  {https://ui.adsabs.harvard.edu/abs/2008ApJ...689..407B} {689, 407}

\bibitem[\protect\citeauthoryear{{Bogdanov}, {Ng}  \& {Kaspi}}{{Bogdanov}
  et~al.}{2014}]{bogdanovetal14}
{Bogdanov} S.,  {Ng} C.~Y.,   {Kaspi} V.~M.,  2014, \mn@doi [\apjl]
  {10.1088/2041-8205/792/2/L36}, \href
  {https://ui.adsabs.harvard.edu/abs/2014ApJ...792L..36B} {792, L36}

\bibitem[\protect\citeauthoryear{{Bogdanov} et~al.,}{{Bogdanov}
  et~al.}{2019}]{bogdanovetal19}
{Bogdanov} S.,  et~al., 2019, \mn@doi [\apjl] {10.3847/2041-8213/ab5968}, \href
  {https://ui.adsabs.harvard.edu/abs/2019ApJ...887L..26B} {887, L26}

\bibitem[\protect\citeauthoryear{{Bonanno}, {Baldo}, {Burgio}  \&
  {Urpin}}{{Bonanno} et~al.}{2014}]{bonannoetal14}
{Bonanno} A.,  {Baldo} M.,  {Burgio} G.~F.,   {Urpin} V.,  2014, \mn@doi [\aap]
  {10.1051/0004-6361/201322514}, \href
  {https://ui.adsabs.harvard.edu/abs/2014A&A...561L...5B} {561, L5}

\bibitem[\protect\citeauthoryear{{Borkowski}, {Reynolds}, {Williams}  \&
  {Petre}}{{Borkowski} et~al.}{2018}]{borkowskietal18}
{Borkowski} K.~J.,  {Reynolds} S.~P.,  {Williams} B.~J.,   {Petre} R.,  2018,
  \mn@doi [\apjl] {10.3847/2041-8213/aaedb5}, \href
  {https://ui.adsabs.harvard.edu/abs/2018ApJ...868L..21B} {868, L21}

\bibitem[\protect\citeauthoryear{{Borkowski}, {Miltich}  \&
  {Reynolds}}{{Borkowski} et~al.}{2020}]{borkowskietal20}
{Borkowski} K.~J.,  {Miltich} W.,   {Reynolds} S.~P.,  2020, \mn@doi [\apjl]
  {10.3847/2041-8213/abcda7}, \href
  {https://ui.adsabs.harvard.edu/abs/2020ApJ...905L..19B} {905, L19}

\bibitem[\protect\citeauthoryear{{Cassam-Chena{\"\i}}, {Decourchelle},
  {Ballet}, {Sauvageot}, {Dubner}  \& {Giacani}}{{Cassam-Chena{\"\i}}
  et~al.}{2004}]{cassamchenaietal04}
{Cassam-Chena{\"\i}} G.,  {Decourchelle} A.,  {Ballet} J.,  {Sauvageot} J.~L.,
  {Dubner} G.,   {Giacani} E.,  2004, \mn@doi [\aap]
  {10.1051/0004-6361:20041154}, \href
  {https://ui.adsabs.harvard.edu/abs/2004A&A...427..199C} {427, 199}

\bibitem[\protect\citeauthoryear{{Chang} \& {Bildsten}}{{Chang} \&
  {Bildsten}}{2003}]{changbildsten03}
{Chang} P.,  {Bildsten} L.,  2003, \mn@doi [\apj] {10.1086/345551}, \href
  {https://ui.adsabs.harvard.edu/abs/2003ApJ...585..464C} {585, 464}

\bibitem[\protect\citeauthoryear{{Chang} \& {Bildsten}}{{Chang} \&
  {Bildsten}}{2004}]{changbildsten04}
{Chang} P.,  {Bildsten} L.,  2004, \mn@doi [\apj] {10.1086/382271}, \href
  {https://ui.adsabs.harvard.edu/abs/2004ApJ...605..830C} {605, 830}

\bibitem[\protect\citeauthoryear{{Chang}, {Bildsten}  \& {Arras}}{{Chang}
  et~al.}{2010}]{changetal10}
{Chang} P.,  {Bildsten} L.,   {Arras} P.,  2010, \mn@doi [\apj]
  {10.1088/0004-637X/723/1/719}, \href
  {https://ui.adsabs.harvard.edu/abs/2010ApJ...723..719C} {723, 719}

\bibitem[\protect\citeauthoryear{{Chevalier}}{{Chevalier}}{1989}]{chevalier89}
{Chevalier} R.~A.,  1989, \mn@doi [\apj] {10.1086/168066}, \href
  {https://ui.adsabs.harvard.edu/abs/1989ApJ...346..847C} {346, 847}

\bibitem[\protect\citeauthoryear{{Davis}}{{Davis}}{2001}]{davis01}
{Davis} J.~E.,  2001, \mn@doi [\apj] {10.1086/323488}, \href
  {https://ui.adsabs.harvard.edu/abs/2001ApJ...562..575D} {562, 575}

\bibitem[\protect\citeauthoryear{{De Luca}}{{De Luca}}{2008}]{deluca08}
{De Luca} A.,  2008, in {Bassa} C.,  {Wang} Z.,  {Cumming} A.,   {Kaspi} V.~M.,
   eds,  American Institute of Physics Conference Series Vol. 983, 40 Years of
  Pulsars: Millisecond Pulsars, Magnetars and More. pp 311--319,
  \mn@doi{10.1063/1.2900173}

\bibitem[\protect\citeauthoryear{{De Luca}}{{De Luca}}{2017}]{deluca17}
{De Luca} A.,  2017, \mn@doi [Journal of Physics Conference Series]
  {10.1088/1742-6596/932/1/012006}, \href
  {https://ui.adsabs.harvard.edu/abs/2017JPhCS.932a2006D} {932, 012006}

\bibitem[\protect\citeauthoryear{{De Luca}, {Mereghetti}, {Caraveo}, {Moroni},
  {Mignani}  \& {Bignami}}{{De Luca} et~al.}{2004}]{delucaetal04}
{De Luca} A.,  {Mereghetti} S.,  {Caraveo} P.~A.,  {Moroni} M.,  {Mignani}
  R.~P.,   {Bignami} G.~F.,  2004, \mn@doi [\aap] {10.1051/0004-6361:20031781},
  \href {https://ui.adsabs.harvard.edu/abs/2004A&A...418..625D} {418, 625}

\bibitem[\protect\citeauthoryear{{DeDeo}, {Psaltis}  \& {Narayan}}{{DeDeo}
  et~al.}{2001}]{dedeoetal01}
{DeDeo} S.,  {Psaltis} D.,   {Narayan} R.,  2001, \mn@doi [\apj]
  {10.1086/322283}, \href
  {https://ui.adsabs.harvard.edu/abs/2001ApJ...559..346D} {559, 346}

\bibitem[\protect\citeauthoryear{{Doroshenko}, {Suleimanov}  \&
  {Santangelo}}{{Doroshenko} et~al.}{2018}]{doroshenkoetal18}
{Doroshenko} V.,  {Suleimanov} V.,   {Santangelo} A.,  2018, \mn@doi [\aap]
  {10.1051/0004-6361/201833271}, \href
  {https://ui.adsabs.harvard.edu/abs/2018A&A...618A..76D} {618, A76}

\bibitem[\protect\citeauthoryear{{Elshamouty}, {Heinke}, {Morsink}, {Bogdanov}
  \& {Stevens}}{{Elshamouty} et~al.}{2016}]{elshamoutyetal16}
{Elshamouty} K.~G.,  {Heinke} C.~O.,  {Morsink} S.~M.,  {Bogdanov} S.,
  {Stevens} A.~L.,  2016, \mn@doi [\apj] {10.3847/0004-637X/826/2/162}, \href
  {https://ui.adsabs.harvard.edu/abs/2016ApJ...826..162E} {826, 162}

\bibitem[\protect\citeauthoryear{{Fesen} et~al.,}{{Fesen}
  et~al.}{2006}]{fesenetal06}
{Fesen} R.~A.,  et~al., 2006, \mn@doi [\apj] {10.1086/504254}, \href
  {https://ui.adsabs.harvard.edu/abs/2006ApJ...645..283F} {645, 283}

\bibitem[\protect\citeauthoryear{{Fesen}, {Kremer}, {Patnaude}  \&
  {Milisavljevic}}{{Fesen} et~al.}{2012}]{fesenetal12}
{Fesen} R.~A.,  {Kremer} R.,  {Patnaude} D.,   {Milisavljevic} D.,  2012,
  \mn@doi [\aj] {10.1088/0004-6256/143/2/27}, \href
  {https://ui.adsabs.harvard.edu/abs/2012AJ....143...27F} {143, 27}

\bibitem[\protect\citeauthoryear{{Fraija}, {Bernal}, {Morales}  \&
  {Negreiros}}{{Fraija} et~al.}{2018}]{fraijaetal18}
{Fraija} N.,  {Bernal} C.~G.,  {Morales} G.,   {Negreiros} R.,  2018, \mn@doi
  [\prd] {10.1103/PhysRevD.98.083012}, \href
  {https://ui.adsabs.harvard.edu/abs/2018PhRvD..98h3012F} {98, 083012}

\bibitem[\protect\citeauthoryear{{Fruscione} et~al.,}{{Fruscione}
  et~al.}{2006}]{fruscioneetal06}
{Fruscione} A.,  et~al., 2006, \mn@doi [Proceedings of SPIE]
  {10.1117/12.671760}, \href
  {https://ui.adsabs.harvard.edu/\#abs/2006SPIE.6270E..1VF} {6270, 62701V}

\bibitem[\protect\citeauthoryear{{Fukui} et~al.,}{{Fukui}
  et~al.}{2003}]{fukuietal03}
{Fukui} Y.,  et~al., 2003, \mn@doi [\pasj] {10.1093/pasj/55.5.L61}, \href
  {https://ui.adsabs.harvard.edu/abs/2003PASJ...55L..61F} {55, L61}

\bibitem[\protect\citeauthoryear{{Gaensler} et~al.,}{{Gaensler}
  et~al.}{2008}]{gaensleretal08}
{Gaensler} B.~M.,  et~al., 2008, \mn@doi [\apjl] {10.1086/589650}, \href
  {https://ui.adsabs.harvard.edu/abs/2008ApJ...680L..37G} {680, L37}

\bibitem[\protect\citeauthoryear{{Gnedin}, {Yakovlev}  \& {Potekhin}}{{Gnedin}
  et~al.}{2001}]{gnedinetal01}
{Gnedin} O.~Y.,  {Yakovlev} D.~G.,   {Potekhin} A.~Y.,  2001, \mn@doi [\mnras]
  {10.1046/j.1365-8711.2001.04359.x}, \href
  {https://ui.adsabs.harvard.edu/abs/2001MNRAS.324..725G} {324, 725}

\bibitem[\protect\citeauthoryear{{Gotthelf} \& {Halpern}}{{Gotthelf} \&
  {Halpern}}{2020}]{gotthelfhalpern20}
{Gotthelf} E.~V.,  {Halpern} J.~P.,  2020, \mn@doi [\apj]
  {10.3847/1538-4357/aba7bc}, \href
  {https://ui.adsabs.harvard.edu/abs/2020ApJ...900..159G} {900, 159}

\bibitem[\protect\citeauthoryear{{Gotthelf}, {Perna}  \& {Halpern}}{{Gotthelf}
  et~al.}{2010}]{gotthelfetal10}
{Gotthelf} E.~V.,  {Perna} R.,   {Halpern} J.~P.,  2010, \mn@doi [\apj]
  {10.1088/0004-637X/724/2/1316}, \href
  {https://ui.adsabs.harvard.edu/abs/2010ApJ...724.1316G} {724, 1316}

\bibitem[\protect\citeauthoryear{{Gotthelf}, {Halpern}  \& {Alford}}{{Gotthelf}
  et~al.}{2013a}]{gotthelfetal13}
{Gotthelf} E.~V.,  {Halpern} J.~P.,   {Alford} J.,  2013a, \mn@doi [\apj]
  {10.1088/0004-637X/765/1/58}, \href
  {https://ui.adsabs.harvard.edu/abs/2013ApJ...765...58G} {765, 58}

\bibitem[\protect\citeauthoryear{{Gotthelf}, {Halpern}, {Allen}  \&
  {Knispel}}{{Gotthelf} et~al.}{2013b}]{gotthelfetal13b}
{Gotthelf} E.~V.,  {Halpern} J.~P.,  {Allen} B.,   {Knispel} B.,  2013b,
  \mn@doi [\apj] {10.1088/0004-637X/773/2/141}, \href
  {https://ui.adsabs.harvard.edu/abs/2013ApJ...773..141G} {773, 141}

\bibitem[\protect\citeauthoryear{{Gourgouliatos}, {Hollerbach}  \&
  {Igoshev}}{{Gourgouliatos} et~al.}{2020}]{gourgouliatosetal20}
{Gourgouliatos} K.~N.,  {Hollerbach} R.,   {Igoshev} A.~P.,  2020, \mn@doi
  [\mnras] {10.1093/mnras/staa1295}, \href
  {https://ui.adsabs.harvard.edu/abs/2020MNRAS.495.1692G} {495, 1692}

\bibitem[\protect\citeauthoryear{{Gusakov}, {Kaminker}, {Yakovlev}  \&
  {Gnedin}}{{Gusakov} et~al.}{2004}]{gusakovetal04}
{Gusakov} M.~E.,  {Kaminker} A.~D.,  {Yakovlev} D.~G.,   {Gnedin} O.~Y.,  2004,
  \mn@doi [\aap] {10.1051/0004-6361:20041006}, \href
  {https://ui.adsabs.harvard.edu/abs/2004A&A...423.1063G} {423, 1063}

\bibitem[\protect\citeauthoryear{{Halpern} \& {Gotthelf}}{{Halpern} \&
  {Gotthelf}}{2010}]{halperngotthelf10}
{Halpern} J.~P.,  {Gotthelf} E.~V.,  2010, \mn@doi [\apj]
  {10.1088/0004-637X/709/1/436}, \href
  {https://ui.adsabs.harvard.edu/abs/2010ApJ...709..436H} {709, 436}

\bibitem[\protect\citeauthoryear{{Hamaguchi}, {Nagata}, {Yanagi}  \&
  {Zheng}}{{Hamaguchi} et~al.}{2018}]{hamaguchietal18}
{Hamaguchi} K.,  {Nagata} N.,  {Yanagi} K.,   {Zheng} J.,  2018, \mn@doi [\prd]
  {10.1103/PhysRevD.98.103015}, \href
  {https://ui.adsabs.harvard.edu/abs/2018PhRvD..98j3015H} {98, 103015}

\bibitem[\protect\citeauthoryear{{Hebbar}, {Heinke}  \& {Ho}}{{Hebbar}
  et~al.}{2020}]{hebbaretal20}
{Hebbar} P.~R.,  {Heinke} C.~O.,   {Ho} W. C.~G.,  2020, \mn@doi [\mnras]
  {10.1093/mnras/stz2570}, \href
  {https://ui.adsabs.harvard.edu/abs/2020MNRAS.491.1585H} {491, 1585}

\bibitem[\protect\citeauthoryear{{Heinke} \& {Ho}}{{Heinke} \&
  {Ho}}{2010}]{heinkeho10}
{Heinke} C.~O.,  {Ho} W. C.~G.,  2010, \mn@doi [\apjl]
  {10.1088/2041-8205/719/2/L167}, \href
  {https://ui.adsabs.harvard.edu/abs/2010ApJ...719L.167H} {719, L167}

\bibitem[\protect\citeauthoryear{{Heinke}, {Rybicki}, {Narayan}  \&
  {Grindlay}}{{Heinke} et~al.}{2006}]{heinkeetal06}
{Heinke} C.~O.,  {Rybicki} G.~B.,  {Narayan} R.,   {Grindlay} J.~E.,  2006,
  \mn@doi [\apj] {10.1086/503701}, \href
  {https://ui.adsabs.harvard.edu/abs/2006ApJ...644.1090H} {644, 1090}

\bibitem[\protect\citeauthoryear{{Ho}}{{Ho}}{2011}]{ho11}
{Ho} W. C.~G.,  2011, \mn@doi [\mnras] {10.1111/j.1365-2966.2011.18576.x},
  \href {https://ui.adsabs.harvard.edu/abs/2011MNRAS.414.2567H} {414, 2567}

\bibitem[\protect\citeauthoryear{{Ho}}{{Ho}}{2013}]{ho13}
{Ho} W. C.~G.,  2013, in {van Leeuwen} J.,  ed.,  International Astronomical
  Union Symposium Vol. 291, Neutron Stars and Pulsars: Challenges and
  Opportunities after 80 years. pp 101--106, \mn@doi{10.1017/S1743921312023289}

\bibitem[\protect\citeauthoryear{{Ho}}{{Ho}}{2014}]{ho14}
{Ho} W. C.~G.,  2014, in {Petit} P.,  {Jardine} M.,   {Spruit} H.~C.,  eds,
  International Astronomical Union Symposium Vol. 302, Magnetic Fields
  throughout Stellar Evolution. pp 435--438, \mn@doi{10.1017/S1743921314002683}

\bibitem[\protect\citeauthoryear{{Ho}}{{Ho}}{2015}]{ho15}
{Ho} W. C.~G.,  2015, \mn@doi [\mnras] {10.1093/mnras/stv1339}, \href
  {https://ui.adsabs.harvard.edu/abs/2015MNRAS.452..845H} {452, 845}

\bibitem[\protect\citeauthoryear{{Ho} \& {Heinke}}{{Ho} \&
  {Heinke}}{2009}]{hoheinke09}
{Ho} W. C.~G.,  {Heinke} C.~O.,  2009, \mn@doi [\nat] {10.1038/nature08525},
  \href {https://ui.adsabs.harvard.edu/abs/2009Natur.462...71H} {462, 71}

\bibitem[\protect\citeauthoryear{{Ho}, {Potekhin}  \& {Chabrier}}{{Ho}
  et~al.}{2008}]{hoetal08}
{Ho} W. C.~G.,  {Potekhin} A.~Y.,   {Chabrier} G.,  2008, \mn@doi [\apjs]
  {10.1086/589238}, \href
  {https://ui.adsabs.harvard.edu/abs/2008ApJS..178..102H} {178, 102}

\bibitem[\protect\citeauthoryear{{Ho}, {Elshamouty}, {Heinke}  \&
  {Potekhin}}{{Ho} et~al.}{2015}]{hoetal15}
{Ho} W. C.~G.,  {Elshamouty} K.~G.,  {Heinke} C.~O.,   {Potekhin} A.~Y.,  2015,
  \mn@doi [\prc] {10.1103/PhysRevC.91.015806}, \href
  {https://ui.adsabs.harvard.edu/abs/2015PhRvC..91a5806H} {91, 015806}

\bibitem[\protect\citeauthoryear{{Katsuda} et~al.,}{{Katsuda}
  et~al.}{2015}]{katsudaetal15}
{Katsuda} S.,  et~al., 2015, \mn@doi [\apj] {10.1088/0004-637X/814/1/29}, \href
  {https://ui.adsabs.harvard.edu/abs/2015ApJ...814...29K} {814, 29}

\bibitem[\protect\citeauthoryear{{Klochkov}, {P{\"u}hlhofer}, {Suleimanov},
  {Simon}, {Werner}  \& {Santangelo}}{{Klochkov} et~al.}{2013}]{klochkovetal13}
{Klochkov} D.,  {P{\"u}hlhofer} G.,  {Suleimanov} V.,  {Simon} S.,  {Werner}
  K.,   {Santangelo} A.,  2013, \mn@doi [\aap] {10.1051/0004-6361/201321740},
  \href {https://ui.adsabs.harvard.edu/abs/2013A&A...556A..41K} {556, A41}

\bibitem[\protect\citeauthoryear{{Klochkov}, {Suleimanov}, {P{\"u}hlhofer},
  {Yakovlev}, {Santangelo}  \& {Werner}}{{Klochkov}
  et~al.}{2015}]{klochkovetal15}
{Klochkov} D.,  {Suleimanov} V.,  {P{\"u}hlhofer} G.,  {Yakovlev} D.~G.,
  {Santangelo} A.,   {Werner} K.,  2015, \mn@doi [\aap]
  {10.1051/0004-6361/201424683}, \href
  {https://ui.adsabs.harvard.edu/abs/2015A&A...573A..53K} {573, A53}

\bibitem[\protect\citeauthoryear{{Klochkov}, {Suleimanov}, {Sasaki}  \&
  {Santangelo}}{{Klochkov} et~al.}{2016}]{klochkovetal16}
{Klochkov} D.,  {Suleimanov} V.,  {Sasaki} M.,   {Santangelo} A.,  2016,
  \mn@doi [\aap] {10.1051/0004-6361/201629208}, \href
  {https://ui.adsabs.harvard.edu/abs/2016A&A...592L..12K} {592, L12}

\bibitem[\protect\citeauthoryear{{Koo}, {Kang}  \& {McClure-Griffiths}}{{Koo}
  et~al.}{2004}]{kooetal04}
{Koo} B.~C.,  {Kang} J.,   {McClure-Griffiths} N.,  2004, in {Camilo} F.,
  {Gaensler} B.~M.,  eds,  International Astronomical Union Symposium Vol. 218,
  Young Neutron Stars and Their Environments. p.~85

\bibitem[\protect\citeauthoryear{{Lamb}, {Boutloukos}, {Van Wassenhove},
  {Chamberlain}, {Lo}, {Clare}, {Yu}  \& {Miller}}{{Lamb}
  et~al.}{2009}]{lambetal09}
{Lamb} F.~K.,  {Boutloukos} S.,  {Van Wassenhove} S.,  {Chamberlain} R.~T.,
  {Lo} K.~H.,  {Clare} A.,  {Yu} W.,   {Miller} M.~C.,  2009, \mn@doi [\apj]
  {10.1088/0004-637X/706/1/417}, \href
  {https://ui.adsabs.harvard.edu/abs/2009ApJ...706..417L} {706, 417}

\bibitem[\protect\citeauthoryear{{Lattimer} \& {Prakash}}{{Lattimer} \&
  {Prakash}}{2001}]{lattimerprakash01}
{Lattimer} J.~M.,  {Prakash} M.,  2001, \mn@doi [\apj] {10.1086/319702}, \href
  {https://ui.adsabs.harvard.edu/abs/2001ApJ...550..426L} {550, 426}

\bibitem[\protect\citeauthoryear{{Lattimer}, {van Riper}, {Prakash}  \&
  {Prakash}}{{Lattimer} et~al.}{1994}]{lattimeretal94}
{Lattimer} J.~M.,  {van Riper} K.~A.,  {Prakash} M.,   {Prakash} M.,  1994,
  \mn@doi [\apj] {10.1086/174025}, \href
  {https://ui.adsabs.harvard.edu/abs/1994ApJ...425..802L} {425, 802}

\bibitem[\protect\citeauthoryear{{Lazendic}, {Slane}, {Gaensler}, {Plucinsky},
  {Hughes}, {Galloway}  \& {Crawford}}{{Lazendic}
  et~al.}{2003}]{lazendicetal03}
{Lazendic} J.~S.,  {Slane} P.~O.,  {Gaensler} B.~M.,  {Plucinsky} P.~P.,
  {Hughes} J.~P.,  {Galloway} D.~K.,   {Crawford} F.,  2003, \mn@doi [\apjl]
  {10.1086/378183}, \href
  {https://ui.adsabs.harvard.edu/abs/2003ApJ...593L..27L} {593, L27}

\bibitem[\protect\citeauthoryear{{Leahy}, {Ranasinghe}  \& {Gelowitz}}{{Leahy}
  et~al.}{2020}]{leahyetal20}
{Leahy} D.~A.,  {Ranasinghe} S.,   {Gelowitz} M.,  2020, \mn@doi [\apjs]
  {10.3847/1538-4365/ab8bd9}, \href
  {https://ui.adsabs.harvard.edu/abs/2020ApJS..248...16L} {248, 16}

\bibitem[\protect\citeauthoryear{{Leinson}}{{Leinson}}{2014}]{leinson14}
{Leinson} L.~B.,  2014, \mn@doi [\jcap] {10.1088/1475-7516/2014/08/031}, \href
  {https://ui.adsabs.harvard.edu/abs/2014JCAP...08..031L} {2014, 031}

\bibitem[\protect\citeauthoryear{{Lovchinsky}, {Slane}, {Gaensler}, {Hughes},
  {Ng}, {Lazendic}, {Gelfand}  \& {Brogan}}{{Lovchinsky}
  et~al.}{2011}]{lovchinskyetal11}
{Lovchinsky} I.,  {Slane} P.,  {Gaensler} B.~M.,  {Hughes} J.~P.,  {Ng} C.~Y.,
  {Lazendic} J.~S.,  {Gelfand} J.~D.,   {Brogan} C.~L.,  2011, \mn@doi [\apj]
  {10.1088/0004-637X/731/1/70}, \href
  {https://ui.adsabs.harvard.edu/abs/2011ApJ...731...70L} {731, 70}

\bibitem[\protect\citeauthoryear{{Luo}, {Ng}, {Ho}, {Bogdanov}, {Kaspi}  \&
  {He}}{{Luo} et~al.}{2015}]{luoetal15}
{Luo} J.,  {Ng} C.~Y.,  {Ho} W.~C.~G.,  {Bogdanov} S.,  {Kaspi} V.~M.,   {He}
  C.,  2015, \mn@doi [\apj] {10.1088/0004-637X/808/2/130}, \href
  {https://ui.adsabs.harvard.edu/abs/2015ApJ...808..130L} {808, 130}

\bibitem[\protect\citeauthoryear{{Mayer}, {Becker}, {Patnaude}, {Winkler}  \&
  {Kraft}}{{Mayer} et~al.}{2020}]{mayeretal20}
{Mayer} M.,  {Becker} W.,  {Patnaude} D.,  {Winkler} P.~F.,   {Kraft} R.,
  2020, \mn@doi [\apj] {10.3847/1538-4357/aba121}, \href
  {https://ui.adsabs.harvard.edu/abs/2020ApJ...899..138M} {899, 138}

\bibitem[\protect\citeauthoryear{{McClure-Griffiths}, {Green}, {Dickey},
  {Gaensler}, {Haynes}  \& {Wieringa}}{{McClure-Griffiths}
  et~al.}{2001}]{mccluregriffithsetal01}
{McClure-Griffiths} N.~M.,  {Green} A.~J.,  {Dickey} J.~M.,  {Gaensler} B.~M.,
  {Haynes} R.~F.,   {Wieringa} M.~H.,  2001, \mn@doi [\apj] {10.1086/320095},
  \href {https://ui.adsabs.harvard.edu/abs/2001ApJ...551..394M} {551, 394}

\bibitem[\protect\citeauthoryear{{Mignani}, {Zaggia}, {de Luca}, {Perna},
  {Bassan}  \& {Caraveo}}{{Mignani} et~al.}{2008}]{mignanietal08}
{Mignani} R.~P.,  {Zaggia} S.,  {de Luca} A.,  {Perna} R.,  {Bassan} N.,
  {Caraveo} P.~A.,  2008, \mn@doi [\aap] {10.1051/0004-6361:20079076}, \href
  {https://ui.adsabs.harvard.edu/abs/2008A&A...484..457M} {484, 457}

\bibitem[\protect\citeauthoryear{{Miller} et~al.,}{{Miller}
  et~al.}{2019}]{milleretal19}
{Miller} M.~C.,  et~al., 2019, \mn@doi [\apjl] {10.3847/2041-8213/ab50c5},
  \href {https://ui.adsabs.harvard.edu/abs/2019ApJ...887L..24M} {887, L24}

\bibitem[\protect\citeauthoryear{{Mori} \& {Ho}}{{Mori} \&
  {Ho}}{2007}]{moriho07}
{Mori} K.,  {Ho} W. C.~G.,  2007, \mn@doi [\mnras]
  {10.1111/j.1365-2966.2007.11663.x}, \href
  {https://ui.adsabs.harvard.edu/abs/2007MNRAS.377..905M} {377, 905}

\bibitem[\protect\citeauthoryear{{Moriguchi}, {Tamura}, {Tawara}, {Sasago},
  {Yamaoka}, {Onishi}  \& {Fukui}}{{Moriguchi} et~al.}{2005}]{moriguchietal05}
{Moriguchi} Y.,  {Tamura} K.,  {Tawara} Y.,  {Sasago} H.,  {Yamaoka} K.,
  {Onishi} T.,   {Fukui} Y.,  2005, \mn@doi [\apj] {10.1086/432653}, \href
  {https://ui.adsabs.harvard.edu/abs/2005ApJ...631..947M} {631, 947}

\bibitem[\protect\citeauthoryear{{Murray}, {Ransom}, {Juda}, {Hwang}  \&
  {Holt}}{{Murray} et~al.}{2002}]{murrayetal02}
{Murray} S.~S.,  {Ransom} S.~M.,  {Juda} M.,  {Hwang} U.,   {Holt} S.~S.,
  2002, \mn@doi [\apj] {10.1086/338224}, \href
  {https://ui.adsabs.harvard.edu/abs/2002ApJ...566.1039M} {566, 1039}

\bibitem[\protect\citeauthoryear{{Negreiros}, {Schramm}  \&
  {Weber}}{{Negreiros} et~al.}{2013}]{negreirosetal13}
{Negreiros} R.,  {Schramm} S.,   {Weber} F.,  2013, \mn@doi [Physics Letters B]
  {10.1016/j.physletb.2012.12.046}, \href
  {https://ui.adsabs.harvard.edu/abs/2013PhLB..718.1176N} {718, 1176}

\bibitem[\protect\citeauthoryear{{Noda}, {Hashimoto}, {Yasutake}, {Maruyama},
  {Tatsumi}  \& {Fujimoto}}{{Noda} et~al.}{2013}]{nodaetal13}
{Noda} T.,  {Hashimoto} M.-a.,  {Yasutake} N.,  {Maruyama} T.,  {Tatsumi} T.,
  {Fujimoto} M.,  2013, \mn@doi [\apj] {10.1088/0004-637X/765/1/1}, \href
  {https://ui.adsabs.harvard.edu/abs/2013ApJ...765....1N} {765, 1}

\bibitem[\protect\citeauthoryear{{Nomoto} \& {Tsuruta}}{{Nomoto} \&
  {Tsuruta}}{1987}]{nomototsuruta87}
{Nomoto} K.,  {Tsuruta} S.,  1987, \mn@doi [\apj] {10.1086/164914}, \href
  {https://ui.adsabs.harvard.edu/abs/1987ApJ...312..711N} {312, 711}

\bibitem[\protect\citeauthoryear{{Page}, {Lattimer}, {Prakash}  \&
  {Steiner}}{{Page} et~al.}{2004}]{pageetal04}
{Page} D.,  {Lattimer} J.~M.,  {Prakash} M.,   {Steiner} A.~W.,  2004, \mn@doi
  [\apjs] {10.1086/424844}, \href
  {https://ui.adsabs.harvard.edu/abs/2004ApJS..155..623P} {155, 623}

\bibitem[\protect\citeauthoryear{{Page}, {Prakash}, {Lattimer}  \&
  {Steiner}}{{Page} et~al.}{2011}]{pageetal11}
{Page} D.,  {Prakash} M.,  {Lattimer} J.~M.,   {Steiner} A.~W.,  2011, \mn@doi
  [\prl] {10.1103/PhysRevLett.106.081101}, \href
  {https://ui.adsabs.harvard.edu/abs/2011PhRvL.106h1101P} {106, 081101}

\bibitem[\protect\citeauthoryear{{Papa} et~al.,}{{Papa}
  et~al.}{2020}]{papaetal20}
{Papa} M.~A.,  et~al., 2020, \mn@doi [\apj] {10.3847/1538-4357/ab92a6}, \href
  {https://ui.adsabs.harvard.edu/abs/2020ApJ...897...22P} {897, 22}

\bibitem[\protect\citeauthoryear{{Park}, {Mori}, {Kargaltsev}, {Slane},
  {Hughes}, {Burrows}, {Garmire}  \& {Pavlov}}{{Park}
  et~al.}{2006}]{parketal06}
{Park} S.,  {Mori} K.,  {Kargaltsev} O.,  {Slane} P.~O.,  {Hughes} J.~P.,
  {Burrows} D.~N.,  {Garmire} G.~P.,   {Pavlov} G.~G.,  2006, \mn@doi [\apjl]
  {10.1086/510366}, \href
  {https://ui.adsabs.harvard.edu/abs/2006ApJ...653L..37P} {653, L37}

\bibitem[\protect\citeauthoryear{{Park}, {Kargaltsev}, {Pavlov}, {Mori},
  {Slane}, {Hughes}, {Burrows}  \& {Garmire}}{{Park} et~al.}{2009}]{parketal09}
{Park} S.,  {Kargaltsev} O.,  {Pavlov} G.~G.,  {Mori} K.,  {Slane} P.~O.,
  {Hughes} J.~P.,  {Burrows} D.~N.,   {Garmire} G.~P.,  2009, \mn@doi [\apj]
  {10.1088/0004-637X/695/1/431}, \href
  {https://ui.adsabs.harvard.edu/abs/2009ApJ...695..431P} {695, 431}

\bibitem[\protect\citeauthoryear{{Pavlov} \& {Luna}}{{Pavlov} \&
  {Luna}}{2009}]{pavlovluna09}
{Pavlov} G.~G.,  {Luna} G.~J.~M.,  2009, \mn@doi [\apj]
  {10.1088/0004-637X/703/1/910}, \href
  {https://ui.adsabs.harvard.edu/abs/2009ApJ...703..910P} {703, 910}

\bibitem[\protect\citeauthoryear{{Pires}}{{Pires}}{2018}]{pires18}
{Pires} A.~M.,  2018, in {Weltevrede} P.,  {Perera} B.~B.~P.,  {Preston} L.~L.,
    {Sanidas} S.,  eds,  International Astronomical Union Symposium Vol. 337,
  Pulsar Astrophysics the Next Fifty Years. pp 112--115,
  \mn@doi{10.1017/S1743921317009590}

\bibitem[\protect\citeauthoryear{{Posselt} \& {Pavlov}}{{Posselt} \&
  {Pavlov}}{2018}]{posseltpavlov18}
{Posselt} B.,  {Pavlov} G.~G.,  2018, \mn@doi [\apj]
  {10.3847/1538-4357/aad7fc}, \href
  {https://ui.adsabs.harvard.edu/abs/2018ApJ...864..135P} {864, 135}

\bibitem[\protect\citeauthoryear{{Posselt}, {Pavlov}, {Suleimanov}  \&
  {Kargaltsev}}{{Posselt} et~al.}{2013}]{posseltetal13}
{Posselt} B.,  {Pavlov} G.~G.,  {Suleimanov} V.,   {Kargaltsev} O.,  2013,
  \mn@doi [\apj] {10.1088/0004-637X/779/2/186}, \href
  {https://ui.adsabs.harvard.edu/abs/2013ApJ...779..186P} {779, 186}

\bibitem[\protect\citeauthoryear{{Potekhin}}{{Potekhin}}{2014}]{potekhin14}
{Potekhin} A.~Y.,  2014, \mn@doi [Physics Uspekhi]
  {10.3367/UFNe.0184.201408a.0793}, \href
  {https://ui.adsabs.harvard.edu/abs/2014PhyU...57..735P} {57, 735}

\bibitem[\protect\citeauthoryear{{Potekhin}, {Fantina}, {Chamel}, {Pearson}  \&
  {Goriely}}{{Potekhin} et~al.}{2013}]{potekhinetal13}
{Potekhin} A.~Y.,  {Fantina} A.~F.,  {Chamel} N.,  {Pearson} J.~M.,   {Goriely}
  S.,  2013, \mn@doi [\aap] {10.1051/0004-6361/201321697}, \href
  {https://ui.adsabs.harvard.edu/abs/2013A&A...560A..48P} {560, A48}

\bibitem[\protect\citeauthoryear{{Potekhin}, {Chabrier}  \& {Ho}}{{Potekhin}
  et~al.}{2014}]{potekhinetal14}
{Potekhin} A.~Y.,  {Chabrier} G.,   {Ho} W.~C.~G.,  2014, \mn@doi [\aap]
  {10.1051/0004-6361/201424619}, \href
  {https://ui.adsabs.harvard.edu/abs/2014A&A...572A..69P} {572, A69}

\bibitem[\protect\citeauthoryear{{Potekhin}, {Pons}  \& {Page}}{{Potekhin}
  et~al.}{2015}]{potekhinetal15}
{Potekhin} A.~Y.,  {Pons} J.~A.,   {Page} D.,  2015, \mn@doi [\ssr]
  {10.1007/s11214-015-0180-9}, \href
  {https://ui.adsabs.harvard.edu/abs/2015SSRv..191..239P} {191, 239}

\bibitem[\protect\citeauthoryear{{Potekhin}, {Zyuzin}, {Yakovlev}, {Beznogov}
  \& {Shibanov}}{{Potekhin} et~al.}{2020}]{potekhinetal20}
{Potekhin} A.~Y.,  {Zyuzin} D.~A.,  {Yakovlev} D.~G.,  {Beznogov} M.~V.,
  {Shibanov} Y.~A.,  2020, \mn@doi [\mnras] {10.1093/mnras/staa1871}, \href
  {https://ui.adsabs.harvard.edu/abs/2020MNRAS.496.5052P} {496, 5052}

\bibitem[\protect\citeauthoryear{{Predehl}, {Costantini}, {Hasinger}  \&
  {Tanaka}}{{Predehl} et~al.}{2003}]{predehletal03}
{Predehl} P.,  {Costantini} E.,  {Hasinger} G.,   {Tanaka} Y.,  2003, \mn@doi
  [Astronomische Nachrichten] {10.1002/asna.200310019}, \href
  {https://ui.adsabs.harvard.edu/abs/2003AN....324...73P} {324, 73}

\bibitem[\protect\citeauthoryear{{Psaltis}, {{\"O}zel}  \& {DeDeo}}{{Psaltis}
  et~al.}{2000}]{psaltisetal00}
{Psaltis} D.,  {{\"O}zel} F.,   {DeDeo} S.,  2000, \mn@doi [\apj]
  {10.1086/317208}, \href
  {https://ui.adsabs.harvard.edu/abs/2000ApJ...544..390P} {544, 390}

\bibitem[\protect\citeauthoryear{{Ransom}}{{Ransom}}{2002}]{ransom02}
{Ransom} S.~M.,  2002, in {Slane} P.~O.,  {Gaensler} B.~M.,  eds,  Astronomical
  Society of the Pacific Conference Series Vol. 271, Neutron Stars in Supernova
  Remnants. p.~361

\bibitem[\protect\citeauthoryear{{Reed}, {Hester}, {Fabian}  \&
  {Winkler}}{{Reed} et~al.}{1995}]{reedetal95}
{Reed} J.~E.,  {Hester} J.~J.,  {Fabian} A.~C.,   {Winkler} P.~F.,  1995,
  \mn@doi [\apj] {10.1086/175308}, \href
  {https://ui.adsabs.harvard.edu/abs/1995ApJ...440..706R} {440, 706}

\bibitem[\protect\citeauthoryear{{Riley} et~al.,}{{Riley}
  et~al.}{2019}]{rileyetal19}
{Riley} T.~E.,  et~al., 2019, \mn@doi [\apjl] {10.3847/2041-8213/ab481c}, \href
  {https://ui.adsabs.harvard.edu/abs/2019ApJ...887L..21R} {887, L21}

\bibitem[\protect\citeauthoryear{{Roger}, {Milne}, {Kesteven}, {Wellington}  \&
  {Haynes}}{{Roger} et~al.}{1988}]{rogeretal88}
{Roger} R.~S.,  {Milne} D.~K.,  {Kesteven} M.~J.,  {Wellington} K.~J.,
  {Haynes} R.~F.,  1988, \mn@doi [\apj] {10.1086/166703}, \href
  {https://ui.adsabs.harvard.edu/abs/1988ApJ...332..940R} {332, 940}

\bibitem[\protect\citeauthoryear{{Sanwal}, {Pavlov}, {Zavlin}  \&
  {Teter}}{{Sanwal} et~al.}{2002}]{sanwaletal02}
{Sanwal} D.,  {Pavlov} G.~G.,  {Zavlin} V.~E.,   {Teter} M.~A.,  2002, \mn@doi
  [\apjl] {10.1086/342368}, \href
  {https://ui.adsabs.harvard.edu/abs/2002ApJ...574L..61S} {574, L61}

\bibitem[\protect\citeauthoryear{{Sedrakian}}{{Sedrakian}}{2013}]{sedrakian13}
{Sedrakian} A.,  2013, \mn@doi [\aap] {10.1051/0004-6361/201321541}, \href
  {https://ui.adsabs.harvard.edu/abs/2013A&A...555L..10S} {555, L10}

\bibitem[\protect\citeauthoryear{{Shabaltas} \& {Lai}}{{Shabaltas} \&
  {Lai}}{2012}]{shabaltaslai12}
{Shabaltas} N.,  {Lai} D.,  2012, \mn@doi [\apj] {10.1088/0004-637X/748/2/148},
  \href {https://ui.adsabs.harvard.edu/abs/2012ApJ...748..148S} {748, 148}

\bibitem[\protect\citeauthoryear{{Shibanov}, {Zavlin}, {Pavlov}  \&
  {Ventura}}{{Shibanov} et~al.}{1992}]{shibanovetal92}
{Shibanov} I.~A.,  {Zavlin} V.~E.,  {Pavlov} G.~G.,   {Ventura} J.,  1992,
  \aap, \href {https://ui.adsabs.harvard.edu/abs/1992A&A...266..313S} {266,
  313}

\bibitem[\protect\citeauthoryear{{Shternin}, {Yakovlev}, {Heinke}, {Ho}  \&
  {Patnaude}}{{Shternin} et~al.}{2011}]{shterninetal11}
{Shternin} P.~S.,  {Yakovlev} D.~G.,  {Heinke} C.~O.,  {Ho} W. C.~G.,
  {Patnaude} D.~J.,  2011, \mn@doi [\mnras] {10.1111/j.1745-3933.2011.01015.x},
  \href {https://ui.adsabs.harvard.edu/abs/2011MNRAS.412L.108S} {412, L108}

\bibitem[\protect\citeauthoryear{{Shternin}, {Ofengeim}, {Ho}, {Heinke},
  {Wijngaarden}  \& {Patnaude}}{{Shternin} et~al.}{2021}]{shterninetal21}
{Shternin} P.~S.,  {Ofengeim} D.~D.,  {Ho} W. C.~G.,  {Heinke} C.~O.,
  {Wijngaarden} M.~J.~P.,   {Patnaude} D.~J.,  2021, \mn@doi [\mnras]
  {10.1093/mnras/stab1695}, \href
  {https://ui.adsabs.harvard.edu/abs/2021MNRAS.tmp.1719S} {506, 709}

\bibitem[\protect\citeauthoryear{{Slane}, {Gaensler}, {Dame}, {Hughes},
  {Plucinsky}  \& {Green}}{{Slane} et~al.}{1999}]{slaneetal99}
{Slane} P.,  {Gaensler} B.~M.,  {Dame} T.~M.,  {Hughes} J.~P.,  {Plucinsky}
  P.~P.,   {Green} A.,  1999, \mn@doi [\apj] {10.1086/307893}, \href
  {https://ui.adsabs.harvard.edu/abs/1999ApJ...525..357S} {525, 357}

\bibitem[\protect\citeauthoryear{{Suleimanov}, {Klochkov}, {Poutanen}  \&
  {Werner}}{{Suleimanov} et~al.}{2017}]{suleimanovetal17}
{Suleimanov} V.~F.,  {Klochkov} D.,  {Poutanen} J.,   {Werner} K.,  2017,
  \mn@doi [\aap] {10.1051/0004-6361/201630028}, \href
  {https://ui.adsabs.harvard.edu/abs/2017A&A...600A..43S} {600, A43}

\bibitem[\protect\citeauthoryear{{Sun}, {Seward}, {Smith}  \& {Slane}}{{Sun}
  et~al.}{2004}]{sunetal04}
{Sun} M.,  {Seward} F.~D.,  {Smith} R.~K.,   {Slane} P.~O.,  2004, \mn@doi
  [\apj] {10.1086/382666}, \href
  {https://ui.adsabs.harvard.edu/abs/2004ApJ...605..742S} {605, 742}

\bibitem[\protect\citeauthoryear{{Tananbaum}}{{Tananbaum}}{1999}]{tananbaum99}
{Tananbaum} H.,  1999, \iaucirc, \href
  {https://ui.adsabs.harvard.edu/abs/1999IAUC.7246....1T} {7246, 1}

\bibitem[\protect\citeauthoryear{{Taranto}, {Burgio}  \& {Schulze}}{{Taranto}
  et~al.}{2016}]{tarantoetal16}
{Taranto} G.,  {Burgio} G.~F.,   {Schulze} H.~J.,  2016, \mn@doi [\mnras]
  {10.1093/mnras/stv2756}, \href
  {https://ui.adsabs.harvard.edu/abs/2016MNRAS.456.1451T} {456, 1451}

\bibitem[\protect\citeauthoryear{{Torii}, {Uchida}, {Hasuike}, {Tsunemi},
  {Yamaguchi}  \& {Shibata}}{{Torii} et~al.}{2006}]{toriietal06}
{Torii} K.,  {Uchida} H.,  {Hasuike} K.,  {Tsunemi} H.,  {Yamaguchi} Y.,
  {Shibata} S.,  2006, \mn@doi [\pasj] {10.1093/pasj/58.1.L11}, \href
  {https://ui.adsabs.harvard.edu/abs/2006PASJ...58L..11T} {58, L11}

\bibitem[\protect\citeauthoryear{{Torres-Forn{\'e}}, {Cerd{\'a}-Dur{\'a}n},
  {Pons}  \& {Font}}{{Torres-Forn{\'e}} et~al.}{2016}]{torresforneetal16}
{Torres-Forn{\'e}} A.,  {Cerd{\'a}-Dur{\'a}n} P.,  {Pons} J.~A.,   {Font}
  J.~A.,  2016, \mn@doi [\mnras] {10.1093/mnras/stv2926}, \href
  {https://ui.adsabs.harvard.edu/abs/2016MNRAS.456.3813T} {456, 3813}

\bibitem[\protect\citeauthoryear{{Tsuji} \& {Uchiyama}}{{Tsuji} \&
  {Uchiyama}}{2016}]{tsujiuchiyama16}
{Tsuji} N.,  {Uchiyama} Y.,  2016, \mn@doi [\pasj] {10.1093/pasj/psw102}, \href
  {https://ui.adsabs.harvard.edu/abs/2016PASJ...68..108T} {68, 108}

\bibitem[\protect\citeauthoryear{{Verner}, {Ferland}, {Korista}  \&
  {Yakovlev}}{{Verner} et~al.}{1996}]{verneretal96}
{Verner} D.~A.,  {Ferland} G.~J.,  {Korista} K.~T.,   {Yakovlev} D.~G.,  1996,
  \mn@doi [\apj] {10.1086/177435}, \href
  {https://ui.adsabs.harvard.edu/abs/1996ApJ...465..487V} {465, 487}

\bibitem[\protect\citeauthoryear{{Vigan{\`o}} \& {Pons}}{{Vigan{\`o}} \&
  {Pons}}{2012}]{viganopons12}
{Vigan{\`o}} D.,  {Pons} J.~A.,  2012, \mn@doi [\mnras]
  {10.1111/j.1365-2966.2012.21679.x}, \href
  {https://ui.adsabs.harvard.edu/abs/2012MNRAS.425.2487V} {425, 2487}

\bibitem[\protect\citeauthoryear{{Wang}, {Qu}  \& {Chen}}{{Wang}
  et~al.}{1997}]{wangetal97}
{Wang} Z.~R.,  {Qu} Q.~Y.,   {Chen} Y.,  1997, \aap, \href
  {https://ui.adsabs.harvard.edu/abs/1997A&A...318L..59W} {318, L59}

\bibitem[\protect\citeauthoryear{{Weinberg}, {Miller}  \& {Lamb}}{{Weinberg}
  et~al.}{2001}]{weinbergetal01}
{Weinberg} N.,  {Miller} M.~C.,   {Lamb} D.~Q.,  2001, \mn@doi [\apj]
  {10.1086/318279}, \href
  {https://ui.adsabs.harvard.edu/abs/2001ApJ...546.1098W} {546, 1098}

\bibitem[\protect\citeauthoryear{{Wijngaarden}, {Ho}, {Chang}, {Heinke},
  {Page}, {Beznogov}  \& {Patnaude}}{{Wijngaarden}
  et~al.}{2019}]{wijngaardenetal19}
{Wijngaarden} M.~J.~P.,  {Ho} W. C.~G.,  {Chang} P.,  {Heinke} C.~O.,  {Page}
  D.,  {Beznogov} M.,   {Patnaude} D.~J.,  2019, \mn@doi [\mnras]
  {10.1093/mnras/stz042}, \href
  {https://ui.adsabs.harvard.edu/abs/2019MNRAS.484..974W} {484, 974}

\bibitem[\protect\citeauthoryear{{Wijngaarden} et~al.,}{{Wijngaarden}
  et~al.}{2020}]{wijngaardenetal20}
{Wijngaarden} M.~J.~P.,  et~al., 2020, \mn@doi [\mnras]
  {10.1093/mnras/staa595}, \href
  {https://ui.adsabs.harvard.edu/abs/2020MNRAS.493.4936W} {493, 4936}

\bibitem[\protect\citeauthoryear{{Williams}, {Hewitt}, {Petre}  \&
  {Temim}}{{Williams} et~al.}{2018}]{williamsetal18}
{Williams} B.~J.,  {Hewitt} J.~W.,  {Petre} R.,   {Temim} T.,  2018, \mn@doi
  [\apj] {10.3847/1538-4357/aaadb6}, \href
  {https://ui.adsabs.harvard.edu/abs/2018ApJ...855..118W} {855, 118}

\bibitem[\protect\citeauthoryear{{Wilms}, {Allen}  \& {McCray}}{{Wilms}
  et~al.}{2000}]{wilmsetal00}
{Wilms} J.,  {Allen} A.,   {McCray} R.,  2000, \mn@doi [\apj] {10.1086/317016},
  \href {http://adsabs.harvard.edu/abs/2000ApJ...542..914W} {542, 914}

\bibitem[\protect\citeauthoryear{{Yang}, {Pi}  \& {Zheng}}{{Yang}
  et~al.}{2011}]{yangetal11}
{Yang} S.-H.,  {Pi} C.-M.,   {Zheng} X.-P.,  2011, \mn@doi [\apjl]
  {10.1088/2041-8205/735/2/L29}, \href
  {https://ui.adsabs.harvard.edu/abs/2011ApJ...735L..29Y} {735, L29}

\bibitem[\protect\citeauthoryear{{Yasumi}, {Nobukawa}, {Nakashima}, {Uchida},
  {Sugawara}, {Tsuru}, {Tanaka}  \& {Koyama}}{{Yasumi}
  et~al.}{2014}]{yasumietal14}
{Yasumi} M.,  {Nobukawa} M.,  {Nakashima} S.,  {Uchida} H.,  {Sugawara} R.,
  {Tsuru} T.~G.,  {Tanaka} T.,   {Koyama} K.,  2014, \mn@doi [\pasj]
  {10.1093/pasj/psu043}, \href
  {https://ui.adsabs.harvard.edu/abs/2014PASJ...66...68Y} {66, 68}

\makeatother
\end{thebibliography}


\appendix

\section{Additional analyses of \Chandra\ and \XMM\ data of \wgaj}
\label{sec:wgajoff}

Here we present analyses of additional spectral data on \wgaj.
These include \Chandra\ observations that are significantly off-axis
and distorted on the ACIS detectors, as discussed in
Sections~\ref{sec:data} and \ref{sec:wgaj}.
Also presented is an analysis where we jointly fit \XMM\ MOS and pn
data, rather than treat each detector separately as done in
Section~\ref{sec:wgaj}.

Table~\ref{tab:spectra_wgaj} shows results of model fits of \Chandra\
spectra at 0.3--7~keV.
We find the pile-up grade migration parameter $\alpha$ to be unconstrained
in fits of the spectra from 2000, and we leave it free to vary since it
has an (small) effect on the uncertainties of other fit parameters.
Note that the blackbody fit parameters, e.g.,
$\Rem\sim1-3\mbox{ km}$, are comparable to those obtained in a joint fit
of the 2000 \Chandra\ spectrum and 2001 \XMM\ spectra \citep{lazendicetal03}.
\citet{lazendicetal03} obtain good fits with multi-component models that
include a power law and either a non-magnetic hydrogen atmosphere
(\texttt{nsspec}) or a magnetic ($B=10^{12}\mbox{ G}$) fully ionized
hydrogen atmosphere (\texttt{nsa}).
Significantly improved single-component model fits are obtained after
considering a wider range of atmosphere model spectra.
Best fits are those that use a partially ionized hydrogen atmosphere model
with $B\ge4\times 10^{12}\mbox{ G}$
(see examples in Table~\ref{tab:spectra_wgaj}), which result in a
$\NH\approx(7-9)\times10^{21}\mbox{ cm$^{-2}$}$ that is somewhat
higher than the SNR absorption $\NH\approx5\times10^{21}\mbox{ cm$^{-2}$}$
near the CCO (see Section~\ref{sec:intro})
and an emission region size $\Rem\ge 4\mbox{ km}$
(for $R=12\mbox{ km}$ and $d=1.3\mbox{ kpc}$).
Atmosphere models with non-magnetic or low magnetic field
($B=10^{10}\mbox{ G}$) hydrogen or non-magnetic carbon also
provide good fits.

\begin{table*}
\centering
\caption{Results of fits to the 2000 and 2014 \Chandra\ spectra of \wgaj\ at 0.3--7~keV.
Along with \texttt{pileup} and \texttt{tbabs}, fits are performed using
model \texttt{bbodyrad}, \texttt{nsatmos}, \texttt{nsx}, or \texttt{nsmaxg}
with model parameters free to vary between 2000 and 2014 fits, unless given otherwise.
Other model parameters are fixed at $M=1.4\,M_\odot$, $R=12\mbox{ km}$,
and $d=1.3\mbox{ kpc}$, unless otherwise noted.
Errors are 1$\sigma$.}
\label{tab:spectra_wgaj}
\begin{tabular}{ccccccc}
\hline
Year & $\alpha$ & $\NH$ & $T$ & $\Rem/R$ & $\fabs$ & $\chi^2$/dof \\
 & & ($10^{21}\mbox{ cm$^{-2}$}$) & ($10^6\mbox{ K}$) & or $\mbox{1.3 kpc}/d$ & ($10^{-13}\mbox{ erg cm$^{-2}$ s$^{-1}$}$) & \\
\hline
\multicolumn{7}{c}{blackbody} \\
2000 & unconstrained & $6.0^{+0.2}_{-0.2}$ & $4.37^{+0.06}_{-0.06}$ & $0.25^{+0.01}_{-0.01}$ & $31.2^{+0.6}_{-0.3}$ & 280.7/230 \\
2014 & $0.24^{+0.03}_{-0.03}$ & $7.4^{+0.5}_{-0.5}$ & $4.13^{+0.10}_{-0.10}$ & $0.088^{+0.006}_{-0.006}$ & $37.1^{+0.2}_{-1.0}$ &         \\
\\
\multicolumn{7}{c}{\texttt{nsatmos} - non-magnetic hydrogen} \\
2000 & unconstrained & $7.4^{+0.2}_{-0.2}$ & $2.63^{+0.05}_{-0.05}$ & $0.97^{+0.05}_{-0.04}$ & $31.7^{+0.3}_{-0.7}$ & 248.5/231 \\
2014 & $0.09^{+0.02}_{-0.02}$ & $9.1^{+0.2}_{-0.2}$ & ---                    & $0.30^{+0.01}_{-0.01}$ & $37.1^{+0.3}_{-0.7}$ &         \\
\\
\multicolumn{7}{c}{\texttt{nsx} - non-magnetic carbon} \\
2000 & unconstrained & $8.1^{+0.1}_{-0.1}$ & $1.62^{+0.06}_{-0.03}$ & $3.7^{+0.3}_{-0.3}$ & $31.9^{+0.1}_{-5.0}$ & 254.4/231 \\
2014 & $0.028^{+0.005}_{-0.006}$ & $8.9^{+0.2}_{-0.2}$ & ---                    & $2.4^{+0.2}_{-0.2}$ & $37.1^{+5.3}_{-2.9}$ &         \\
\\
\multicolumn{7}{c}{\texttt{nsmaxg} - hydrogen at $10^{10}\mbox{ G}$ ($R=10\mbox{ km}$)} \\
2000 & unconstrained & $7.6^{+0.2}_{-0.2}$ & $2.90^{+0.05}_{-0.05}$ & $1.03^{+0.05}_{-0.05}$ & $31.7^{+0.3}_{-0.8}$ & 250.0/231 \\
2014 & $0.08^{+0.02}_{-0.02}$ & $8.7^{+0.3}_{-0.2}$ & ---                    & $0.32^{+0.01}_{-0.01}$ & $36.9^{+0.5}_{-0.7}$ &         \\
\\
\multicolumn{7}{c}{\texttt{nsmaxg} - hydrogen at $10^{13}\mbox{ G}$} \\
2000 & unconstrained & $8.2^{+0.2}_{-0.1}$ & $2.67^{+0.05}_{-0.06}$ & $1.03^{+0.07}_{-0.05}$ & $31.7^{+0.2}_{-0.6}$ & 241.7/231 \\
2014 & $0.09^{+0.02}_{-0.01}$ & $9.3^{+0.3}_{-0.3}$ & ---                    & $0.32^{+0.02}_{-0.01}$ & $37.0^{+0.3}_{-0.5}$ &         \\
\hline
\end{tabular}
\end{table*}

\begin{figure}
\begin{center}
\includegraphics[width=0.45\textwidth]{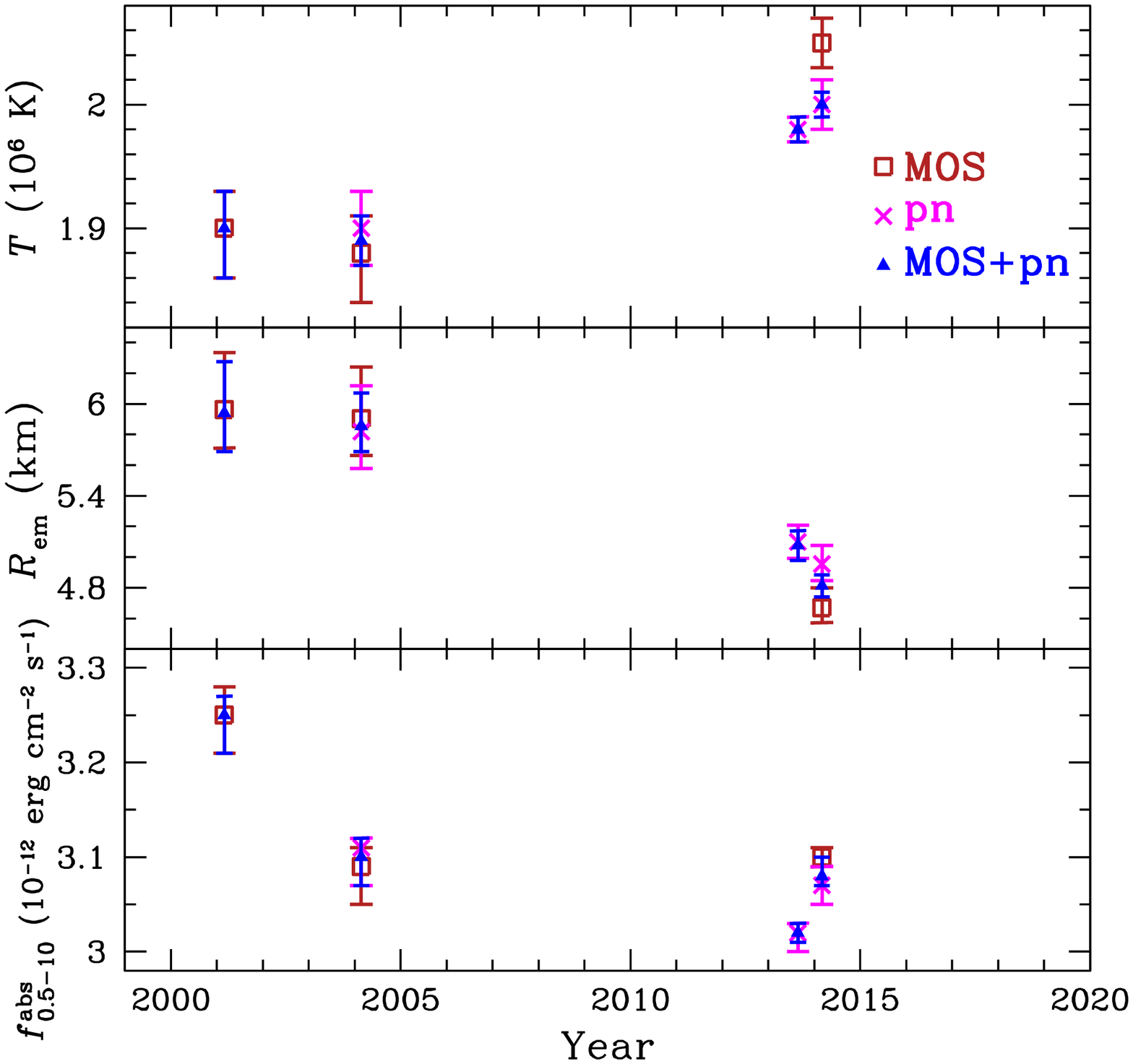}
\caption{
Surface temperature $T$ (top), emission radius $\Rem$ (middle),
and absorbed 0.5--10~keV flux $\fabs$ (bottom) of \wgaj\
from non-magnetic carbon atmosphere model fits to \XMM\ MOS (squares)
and pn (crosses) spectra independently (see Tables~\ref{tab:xmmmos}
and \ref{tab:xmmpn})
and fits to MOS+pn (triangles) spectra jointly (see Table~\ref{tab:xmmjoint}).
Errors are 1$\sigma$.
}
\label{fig:wgajjoint}
\end{center}
\end{figure}

Table~\ref{tab:xmmjoint} shows results of joint model fits of \XMM\ MOS
spectra at 0.2--7~keV and pn spectra at 0.3--10~keV, respectively.
Note that, in these joint fits, an extra fit parameter is introduced
to account for a difference in relative normalization between MOS
and pn spectra; this parameter is allowed to vary in fits of the 2004
and 2014 MOS and pn data, while it is fixed to 1 for the 2001 MOS and
2013 pn data.
In Figure~\ref{fig:wgajjoint}, we show the evolution of the MOS and pn
inferred temperature $T$, emission radius $\Rem$, and absorbed flux
$\fabs$ for non-magnetic carbon atmosphere model fits.
These results are very similar to those obtained in Section~\ref{sec:wgaj}
where MOS and pn data are fit independently.

\begin{table*}
\centering
\caption{Results of joint fits to the 2001, 2004, and 2014 \XMM\
EPIC-MOS spectra at 0.2--7~keV and 2004, 2013, and 2014 \XMM\ EPIC-pn
spectra at 0.3--10~keV of \wgaj.
Along with \texttt{tbabs} and a parameter to account for MOS and pn
differences, fits are performed using
model \texttt{bbodyrad}, \texttt{nsatmos}, \texttt{nsx}, or \texttt{nsmaxg}
with model parameters free to vary between fits, unless given otherwise.
Other model parameters are fixed at $M=1.4\,M_\odot$, $R=12\mbox{ km}$,
and $d=1.3\mbox{ kpc}$, unless otherwise noted.
Absorbed 0.5--10~keV flux $\fabs$ is in $10^{-13}\mbox{ erg cm$^{-2}$ s$^{-1}$}$.
Errors are 1$\sigma$.}
\label{tab:xmmjoint}
\begin{tabular}{ccccccc}
\hline
Year & $\NH$ & $T$ & $\Rem/R$ & $\fabs$ & MOS/pn & $\chi^2$/dof \\
& ($10^{21}\mbox{ cm$^{-2}$}$) & ($10^6\mbox{ K}$) & or $\mbox{1.3 kpc}/d$ & & & \\
\hline
\multicolumn{7}{c}{blackbody} \\
2001 & $4.51^{+0.19}_{-0.18}$ & $4.75^{+0.05}_{-0.05}$ & $0.0442^{+0.0014}_{-0.0013}$ & $32.0^{+0.4}_{-0.3}$ & --- & $3421/2188$ \\
2004 & $4.58^{+0.11}_{-0.11}$ & $4.73^{+0.04}_{-0.04}$ & $0.0437^{+0.0009}_{-0.0008}$ & $30.5^{+0.2}_{-0.4}$ & $1.00^{+0.01}_{-0.01}$ &  \\
2013 & $4.68^{+0.05}_{-0.05}$ & $4.82^{+0.02}_{-0.02}$ & $0.0414^{+0.0004}_{-0.0004}$ & $29.7^{+0.1}_{-0.1}$ & --- &  \\
2014 & $4.57^{+0.05}_{-0.05}$ & $4.90^{+0.02}_{-0.01}$ & $0.0402^{+0.0003}_{-0.0003}$ & $30.4^{+0.2}_{-0.1}$ & $1.000^{+0.005}_{-0.005}$ &  \\
\\
\multicolumn{7}{c}{\texttt{nsatmos} - non-magnetic hydrogen} \\
2001 & $6.01^{+0.04}_{-0.04}$ & $2.93^{+0.03}_{-0.04}$ & $0.161^{+0.005}_{-0.003}$ & $32.6^{+0.2}_{-0.4}$ & --- & $2396/2191$\\
2004 & ---                    & $2.93^{+0.02}_{-0.02}$ & $0.158^{+0.003}_{-0.002}$ & $31.0^{+0.3}_{-0.3}$ & $1.00^{+0.01}_{-0.01}$ & \\
2013 & ---                    & $3.01^{+0.01}_{-0.01}$ & $0.146^{+0.002}_{-0.001}$ & $30.2^{+0.1}_{-0.1}$ & --- & \\
2014 & ---                    & $3.06^{+0.01}_{-0.01}$ & $0.142^{+0.001}_{-0.001}$ & $30.9^{+0.1}_{-0.2}$ & $1.000^{+0.005}_{-0.005}$ & \\
\\
\multicolumn{7}{c}{\texttt{nsx} - non-magnetic carbon} \\
2001 & $7.05^{+0.04}_{-0.04}$ & $1.90^{+0.03}_{-0.04}$ & $0.495^{+0.028}_{-0.021}$ & $32.5^{+0.2}_{-0.4}$ & --- & $2354/2191$ \\
2004 & ---                    & $1.89^{+0.02}_{-0.02}$ & $0.488^{+0.018}_{-0.014}$ & $31.0^{+0.2}_{-0.3}$ & $1.00^{+0.01}_{-0.01}$ & \\
2013 & ---                    & $1.98^{+0.01}_{-0.01}$ & $0.423^{+0.008}_{-0.008}$ & $30.2^{+0.1}_{-0.1}$ & --- & \\
2014 & ---                    & $2.00^{+0.01}_{-0.01}$ & $0.401^{+0.006}_{-0.006}$ & $30.8^{+0.2}_{-0.1}$ & $0.999^{+0.005}_{-0.005}$ & \\
\\
\multicolumn{7}{c}{\texttt{nsmaxg} - hydrogen at $3\times10^{13}\mbox{ G}$ ($R=10\mbox{ km}$)} \\
2001 & $6.73^{+0.03}_{-0.05}$ & $3.09^{+0.03}_{-0.03}$ & $0.196^{+0.005}_{-0.005}$ & $32.6^{+0.3}_{-0.3}$ & --- & $2458/2191$ \\
2004 & ---                    & $3.08^{+0.02}_{-0.02}$ & $0.193^{+0.003}_{-0.003}$ & $31.1^{+0.3}_{-0.3}$ & $1.00^{+0.01}_{-0.01}$ & \\
2013 & ---                    & $3.17^{+0.02}_{-0.01}$ & $0.177^{+0.001}_{-0.002}$ & $30.2^{+0.2}_{-0.1}$ & --- & \\
2014 & ---                    & $3.21^{+0.02}_{-0.01}$ & $0.173^{+0.001}_{-0.002}$ & $30.9^{+0.1}_{-0.1}$ & $0.998^{+0.005}_{-0.005}$ & \\
\hline
\end{tabular}
\end{table*}

\bsp	
\label{lastpage}
\end{document}